%% file: main.tex
\documentclass[a4paper, 12pt]{article}
\usepackage{natbib}

\usepackage[english]{babel}
\usepackage[utf8x]{inputenc}
\usepackage[T1]{fontenc}
\usepackage{listings}
\usepackage{fixmath}
\usepackage{MnSymbol}
\usepackage{float}

\usepackage{tikz}
\usepackage{graphicx}
\usetikzlibrary{positioning}

\usepackage[a4paper,top=2.5cm,bottom=2.5cm,left=3cm,right=3cm,marginparwidth=2.cm]{geometry}
\usepackage{setspace}
\usepackage{caption}
\usepackage{amsmath}
\numberwithin{equation}{section}
\usepackage{graphicx}
\usepackage[colorlinks=true, allcolors=blue]{hyperref}

\usepackage{titlesec}
\titlelabel{\thetitle.\;}
\usepackage{lipsum}
\usepackage{enumitem,calc}
\usepackage{amsmath}
\usepackage{bbold}
\usepackage{mathtools}

\newcommand{\var}{{\mbox{Var}}}
\newcommand{\E}{\mbox{E}}

\newcommand{\bmu}{{\mbox{\boldmath $\mu$}}} 
\newcommand{\balpha}{{\mbox{\boldmath $\alpha$}}}

\newcommand{\bbeta}{{\mbox{\boldmath $\beta$}}}

\newcommand{\bkappa}{{\mbox{\boldmath $\kappa$}}}

\newcommand{\bSigma}{{\mathbf{\Sigma}}}

\newcommand{\be}{\mathbold{e}}
\newcommand{\bm}{\mathbold{m}}
\newcommand{\bv}{\mathbold{v}}
\newcommand{\bw}{\mathbold{w}}
\newcommand{\by}{\mathbold{y}}
\newcommand{\bF}{\mathbf{F}}
\newcommand{\bG}{\mathbf{G}}
\newcommand{\bK}{\mathbf{K}}
\newcommand{\bR}{\mathbf{R}}
\newcommand{\bW}{\mathbf{W}}
\newcommand{\bV}{\mathbf{V}}
\newcommand{\bI}{\mathbf{I}}
\newcommand{\bJ}{\mathbold{J}}
\newcommand{\bL}{\mathbf{L}}
\newcommand{\bzero}{{\mbox{\boldmath $0$}}}
\title{Bayesian Poisson Mortality Projections with Incomplete Data}
\date{}
\author{Rui Gong$^1$, Xiaoqian Sun$^1$, Leping Liu$^2$\thanks{Part of his research has been supported by National Science Foundation of China (NSFC) under Grant \#71771163.}, Yu-Bo Wang$^{1}$\thanks{Corresponding author: yubow@clemson.edu}\\ 
		$^1$School of Mathematical and Statistical Sciences, \\Clemson University, SC, USA\\

		$^2$Department of Statistics, Tianjin University of \\Finance and Economics, Tianjin, China}

\begin{document}
\maketitle


\newpage
\centerline{Abstract}
       The missing data problem pervasively exists in statistical applications. Even as simple as the count data in mortality projections, it may not be available for certain age-and-year groups due to the budget limitations or difficulties in tracing research units, resulting in the follow-up estimation and prediction inaccuracies. To circumvent this data-driven challenge, we extend the Poisson log-normal Lee-Carter model to accommodate a more flexible time structure, and develop the new sampling algorithm that improves the MCMC convergence when dealing with incomplete mortality data. 
       Via the overdispersion term and Gibbs sampler, the extended model can be re-written as the dynamic linear model so that both Kalman and sequential Kalman filters can be incorporated into the sampling scheme. 
       Additionally, our meticulous prior settings can avoid the re-scaling step in each MCMC iteration, and allow model selection simultaneously conducted with estimation and prediction. 
       The proposed method is applied to the mortality data of Chinese males during the period 1995-2016 to yield mortality rate forecasts for 2017-2039. The results are comparable to those based on the imputed data set, suggesting that our approach could handle incomplete data well.


\doublespacing
\input{Section1.tex}

\input{Section2.tex}

\input{Section3.tex}
\input{Section4.tex}

\input{Section5.tex}

\input{Section6.tex}

\bibliographystyle{apalike}


\bibliographystyle{apalike}

\end{document}

%% file: Section1.tex
\section{Introduction} \label{introduction}
 \hspace{12pt}  
 Population structure always plays an important role in the socio-economic policy decisions. For example, the rapid change of life expectancy keeps affecting and altering the current retirement systems, healthy cares, and annuities. It also has a great impact on government allocation of funds and intergenerational resource transfer \citep{11,111,1111}.  
 Therefore, mortality projections that provide a glimpse of future population structures have become a focus in modern demographics.  
 
 

 The Lee-Carter (LC) model \citep{Lee}, which was originally developed for the U.S. mortality projections, has now been widely implemented in different types of mortality data due to its seminal and easily interpreted log-bilinear structure. In the proposed framework, the observed log mortality rates are first decomposed into the age and time related effects via the singular value decomposition (SVD), and then based on the estimated time effects, a separate time series model is fitted to obtain the future trajectory of mortality. Clearly, as the output of this two-stage analysis, the prediction intervals only preserve uncertainty from the second model and are underestimated. 
In the light of this, \citet{Claudia} develops the Bayesian LC model to properly incorporate all sources of variability into mortality projections. The proposed method also improves the Markov chain Monte Carlo (MCMC) convergence when missing data exists by introducing the Kalman filter into the sampling scheme. 
Following this line, \citet{Johnny} further develop the sequential Kalman filter to grant the investigators flexibility in handling missing data. Specifically, when missing mechanisms are not clear or heterogeneity is suggested between observations and missing ones, the modified algorithm can directly handle the incomplete data set without imputations.
 
 
Besides modeling on the observed mortality rates, it is reasonable to assume the number of deaths following a Poisson distribution with mean equal to the population size multiplied by the underlying true mortality rate, which is hierarchically controlled by the age and time effects. Driven by this thought, \citet{Brouhns2,Brouhns} utilize both death counts and exposures at risk in the model fitting and develop the Poisson LC (PLC) model. Although this Poisson framework has a limitation in handling the count data with overdispersion, \citet{Jackie} propose the Poisson log-normal LC (PLNLC) model to address this issue. Other related efforts to relax this restriction can be found in \citet{Delwarde}, \citet{Renshaw}, and \citet{Li}. 

In this work, we find a gap between the applications of Poisson Lee–Carter framework and incomplete mortality data. To allow incomplete data directly handled under the Poisson framework, we extend the PLNLC model to a more general time structure, and through the Gaussian overdispersion term and an MCMC sample of the mortality rates, we successfully combine the sequential Kalman filter into the Gibbs sampler to improve the MCMC algorithm. 
In some sense, the proposed method can be viewed as a twins work of \citet{Johnny}. 
Also, inspired by \citet{Jackie} and \citet{Zhen}, this model features with two meticulous prior settings: first, we adopt the priors of age effects subjected to the constraints to skip the re-scaling step in each iteration; secondly, we consider the dirac spike \citep{Wagner} together with the time structure model so that the competing nested structures can be compared and selected in one single analysis. 

 The remainder of this paper is organized as follows. In Section 2, we review the recent developments of the LC model. Section 3 provides the details of the proposed model and its prior specifications. Section 4 develops the MCMC sampling algorithm for both complete and incomplete data. In Section 5, the proposed method is applied to the incomplete mortality data set of Chinese males in the years 1995-2016. For the comparison purposes, the results based on SVD-imputed complete data are also provided. At last, we conclude with a discussion in Section 5.

%% file: Section2.tex
\section{Review of Recent Mortality Models}
Suppose the mortality data records the death tolls $D_{x,t}$ for $M$ age groups across $N$ years; i.e., $x\in \Theta_{\text{age}}=\{1,2,\dots,M\}$ and $t\in\Theta_{\text{time}}=\{1,2,\dots,N\}$, and let $E_{x,t}$ denote the corresponding population size at the risk. \citet{Claudia} formalized the Bayesian version of LC model \citep{Lee} to properly present the uncertainty of mortality projections, and incorporated multiple imputations \citep{Rubin} to address the missing data problems. Based on the observed mortality rates $m_{x,t}=D_{x,t}/E_{x,t}$, the joint model is given by 
\begin{align}
    &\log(m_{x,t}) = \alpha_x +\beta_x\kappa_t + \varepsilon_{x,t},\label{eq2.1}\\
    &\kappa_t=\theta+\kappa_{t-1}+\omega_t,\label{eq2.2}
\end{align}
where $\alpha_x$ is the age-specific intercept denoting the average log mortality rate at age $x$ over $N$ years under the constraints $\sum_{x\in \Theta_{\text{age}}}\beta_x=1$ and $\sum_{t\in \Theta_{\text{time}}}\kappa_t=0$,  $\beta_x$ is the measured sensitivity of each age group to the overall trend of mortality that is captured by $\kappa_t$ in a random walk with drift model (\ref{eq2.2}), $\theta$ is the drift term, and $\varepsilon_{x,t} \stackrel{iid}{\sim}N(0,\sigma_{\varepsilon}^2)$ and $\omega_t \stackrel{iid}{\sim}N(0,\sigma_{\omega}^2)$. Unlike the original LC model, the error terms in (\ref{eq2.1}) have Gaussian specifications. As a result, all full conditionals for the Gibbs sampling are analytically tractable if conjugate priors are assigned, and Kalman filter \citep{Harvey} can be implemented to improve efficiency of MCMC sampling. Specifically, Kalman filter consists of the filtering and smoothing processes, which rely on the up-to-now (from the $1^{st}$ to $t^{th}$ year) and beyond-time-t (from the $(t+1)^{th}$ to $N^{th}$ year) information, respectively, to form the full conditional distribution of $\kappa_{t}$. 

As pointed out by \citet{Johnny}, however, the aforementioned method may fail to obtain a convergent MCMC sample if the initial values for a Gibbs sampler are not chosen carefully, especially when the case contains a large proportion of missing data. To circumvent this difficulty, \citet{Johnny} proposed the new procedure to generate these values that are sufficiently close to the stationary state for the Gibbs sampler. They also claimed that only missing values appearing in a sporadic manner require imputations since those shown as blocks can be viewed as missing completely at random, and leaving them blank does not affect the follow-up analyses. Accordingly, they developed the abridged multiple imputation, and based on this partially imputed data set, adopted the sequential Kalman filter \citep{Durbin} to update $\kappa_t$'s in each iteration. The details of Kalman and sequential Kalman filters are hold till Section 4 about our sampling algorithms.

Another thread of derivations of the classic LC model can be traced back to  \citet{Brouhns2,Brouhns}, where the observed death count $D_{x,t}$ is assumed following a Poisson distribution with mean equal to $E_{x,t}$ times unknown mortality rate $\mu_{x,t}\coloneqq\exp(\alpha_x +\beta_x\kappa_t)$. Proceeding in this fashion, the proposed PLC model distinguishes the cases with the same observed mortality rate but different exposure sizes so that utilizing more information from data. \citet{Jackie} further developed the PLNLC model to accommodate overdispersion, commonly present in the Poisson applications,
by introducing the random effects $\varepsilon_{x,t} \stackrel{iid}{\sim}N(0,\sigma_{\varepsilon}^2)$ to the PLC model as follows  
\begin{align}
\label{eq2.3}
    & D_{x,t}\mid \mu_{x,t}\sim\text{Poisson} (E_{x,t}\mu_{x,t}),\nonumber\\
    & \log(\mu_{x,t}) = \alpha_x +\beta_x\kappa_t + \varepsilon_{x,t}.
\end{align}
Consequently, 
\begin{align*}
    \var(D_{x,t}\mid \mu_{x,t})=&\E[\var(D_{x,t}\mid\mu_{x,t},\varepsilon_{x,t})]+\var[\E(D_{x,t}\mid\mu_{x,t},\varepsilon_{x,t})]\\
=&\E(D_{x,t}\mid \mu_{x,t})\times\{1+\E(D_{x,t}\mid \mu_{x,t})\times[\exp(\sigma_\varepsilon^2)-1]\}\geq \E(D_{x,t}\mid \mu_{x,t}),
\end{align*}
 and thus (\ref{eq2.3}) can handle count data having greater variability than its expectation.
\citet{Jackie} also adopted the new pair of constraints $\sum_{x\in \Theta_{\text{age}}}\alpha_x=1$ and $\kappa_1=0$ to directly embed in the prior specifications of $\beta_x$'s and $\kappa_t$'s. Different from the MCMC sampling algorithms in \citet{Claudia} and \citet{Johnny} requiring the re-scaling adjustments in each iteration to meet the constraints (this adjustment seems lacking of theoretical justifications), the proposed algorithm simplifies the sampling procedure for $\beta_x$'s and $\kappa_t$'s, and  avoid the potential issue in ergodic conditions. 

Motivated by \citet{Jackie} and \citet{Johnny}, we propose the new Bayesian approach to address the mortality projections complicated with missing data under the Poisson framework. Specifically, we follow the PLNLC model but provide a new insight of $\varepsilon_{x,t}$ in (\ref{eq2.3}): not only accommodating overdispersion but also having $\log(\mu_{x,t})\stackrel{iid}{\sim}N(\alpha_x+\beta_x\kappa_t,\sigma_{\varepsilon}^2)$. With such a Gaussian expression, Kalman and sequential Kalman filters can now be implemented to improve efficiency of MCMC sampling given that $\log(\mu_{x,t})$ is available. To this end, we simply involve $\log(\mu_{x,t})$ in MCMC sampling, and state space form of PLNLC is hold for the full conditional distribution of $\kappa_t$.
We also extend (\ref{eq2.2}) to a more general setting to fit mortality data with more varieties of time trends; meanwhile, inspired by \citet{Zhen}, the dirac spike prior \citep{Wagner} is used to conduct model selection on the time structure simultaneously with estimation. Lastly, to embed the constraints in the prior specifications as \citet{Jackie} and keep the state space form of PLNLC, we alter the constraints as $\sum_{x\in \Theta_{\text{age}}}\alpha_x=0$ and $\sum_{x\in \Theta_{\text{age}}}\beta_x=1$.

%% file: Section3.tex
\section{The Proposed Model and its Properties} 
Let $\theta_1$ stand for the drift term $\theta$ in (\ref{eq2.2}) while $\theta_2$ is the additive slope of a random walk with drift model, our proposed joint model is given by
\begin{align}
    & D_{x,t}\mid \mu_{x,t} \sim \text{Poisson}(E_{x,t}\mu_{x,t}),     \label{eq3.1}  \\ 
    & \log(\mu_{x,t})=\alpha_x + \beta_x\kappa_t+\varepsilon_{x,t}, \label{eq3.2} \\
    & \kappa_t = \kappa_{t-1} +\theta_1 +\theta_2t +\omega_t, \label{eq3.3} 
\end{align}
where $\kappa_0\sim N(\mu_{\kappa_{0}}\,,\,\sigma_{\kappa_{0}}^2)$, and $\mu_{\kappa_{0}}$ and $\sigma_{\kappa_{0}}^2$ are pre-specified values. Although it appears that this model is the same as PLNLC by \citet{Jackie} except that a more general setting of the time structure is considered, we highlight our differences and novelties in the following three subsections. Throughout the paper, we use the superscript $T$ as the transposition of a vector or a matrix. We also introduce the notations $\bzero_n$, $\bJ_n$, and $\bI_n$ to represent a vector of zeros with size $n$, a vector of ones with size n, and an identity matrix with size n, respectively. For convenience, if no specifications on bounds, $\sum_x=\sum_{x\in\Theta_{\text{age}}}$ and $\sum_t=\sum_{t\in\Theta_{\text{time}}}$. 

\subsection{Conditional State Space Form of PLNLC}
First, with the Gaussian specifications of $\varepsilon_{x,t}$ and $\omega_t$, the PLNLC model is readily expressed in a state space form, where (\ref{eq3.2}) and (\ref{eq3.3}) separately serve as observation and state equations in the Kalman filter. However, due to the unobservable $\log(\mu_{x,t})$, the linear quadratic estimation for $\kappa_t$ is still hindered from use. To circumvent this difficulty, we let the latent variable $\log(\mu_{x,t})$ involved in a Gibbs sampler. Specifically, based on  (\ref{eq3.1}) and (\ref{eq3.2}), the full conditional $\pi\left(\log(\mu_{x,t})\mid D_{x,t}, \alpha_x, \beta_x, \kappa_t,\sigma_{\varepsilon}^2\right)$ is proportional to an analogously Gaussian kernel as follows
\begin{align}
    \mu_{x,t}^{D_{x,t}}\exp\left[-E_{x,t}\mu_{x,t}-\frac{1}{2\sigma_{\varepsilon}^2}\left(\log(\mu_{x,t})-\alpha_x - \beta_x\kappa_t\right)^2\right].\label{logmu}
\end{align}
Accordingly, assuming that $\log^{(i)}(\mu_{x,t})$, the $i^{th} $ iteration of $\log(\mu_{x,t})$, is available, we propose $\log'(\mu_{x,t})\sim N(\log^{(i)}(\mu_{x,t}), \sigma_{x,t}^2)$ with a pre-specified $\sigma_{x,t}^2$ to update    
\begin{align*}
\log^{(i+1)}(\mu_{x,t}) =
  \begin{cases}
    \log'(\mu_{x,t})       & \quad \text{if } u \leq \phi(\log'(\mu_{x,t}),\log^{(i)}(\mu_{x,t})),\\
    \log^{(i)}(\mu_{x,t})  & \quad \text{otherwise},
  \end{cases}
\end{align*}
where  $u\sim\text{Uniform}(0,1)$ and 
\begin{align*}
    &\phi \left(\log'(\mu_{x,t}),\log^{(i)}(\mu_{x,t})\right)\\ 
    = &\min \; \left(1,\; \left(\frac{\mu'_{x,t}}{\mu^{(i)}_{x,t}}\right)^{D_{x,t}}\exp\left\{-E_{x,t}(\mu'_{x,t}-\mu^{(i)}_{x,t})-\frac{1}{2\sigma_{\varepsilon}^2}\log\left(\frac{\mu'_{x,t}}{\mu^{(i)}_{x,t}}\right)\left[\log(\mu'_{x,t}\mu^{(i)}_{x,t})-2\alpha_x - 2\beta_x\kappa_t\right]\right\}\right).
\end{align*}
Once $\log(\mu_{x,t})$ is attained in the Metropolis-within-Gibbs algorithm, the Kalman and sequential Kalman filters can be implemented. As shown that this realization is contingent on the Gibbs sampler, we name this feature as the conditional state space form of PLNLC. 

It is also worth mentioning that the chosen values of $\sigma_{x,t}^2$'s affect the acceptance rates of  $\log'(\mu_{x,t})$'s. To ensure the acceptance rates between 0.15 and 0.5, the interval suggested by \citet{Roberts}, we adopt the trial and error search procedure proposed by \citet{Jackie} to determine $\sigma_{x,t}^2$'s. In particular, we start with an initial value 0.01 for all $\sigma_{x,t}^2$'s, and evaluate their acceptance rates every 100 iterations. If any rates are above 0.5 (or below 0.15), we double (or halve) the values of corresponding $\sigma_{x,t}^2$'s in the next 100-iteration cycle; otherwise, keep them the same. We repeat this searching procedure till the $20^{th}$ cycle.  




\subsection{Model Selection on Time Structure}
As previously mentioned,  (\ref{eq3.3}) presents a more flexible setting of time structure, and can reduce back to a random walk with drift model when $\theta_2=0$. To allow the data to reflect its own structure, that is, $\theta_2$ is zero or non-zero, and avoid additional model selection procedure, we propose the dirac spike prior on $\theta_2$ as follows  
\begin{align*}
    &\theta_2\sim z N\left(0,\zeta\right)+\left(1-z\right)\pi,\\
    &z \sim \text{Bernoulli}(p_0),
\end{align*}
where $z$ is a binary indicator with $z=1$ favoring the full model of time structure while $z=0$ favors the reduced one, $\zeta$ is a random scalar controlling the variation of non-zero $\theta_2$,  $\pi$ is a point mass at zero, and $p_0$ is a prior belief of probability that $\theta_2$ is non-zero. 
Under such a setting, a Gibbs sample of $z$ is updated with the conditional posterior probability 
\begin{align*}
\begin{split}
&\tilde{p} = 1 - \frac{1-p_0}{1-p_0 +p_0\sqrt{\frac{\sigma_{\omega}^2}{\zeta\sum_t{t^2} + 
\sigma_{\omega}^2}}\exp\left\{
\frac{\left[\sum_t{t(\kappa_t-\kappa_{t-1}-\theta_1})\right]^2\zeta}{2\sigma_{\omega}^2(\sum_t{t^2}\zeta+\sigma_{\omega}^2)}\right\}}.
\end{split}  
\end{align*}
We then can find out the best fitted time structure by simply taking the average of an MCMC sample of $z$, i.e., if the value is greater than 0.5, the more complicated structure is selected; otherwise, a random walk with drift model.

\subsection{Prior Specifications Subjected to the Constraints}
As the LC model becomes a benchmark stochastic model for mortality data, some potential issues and limitations regarding to the constraint $\sum_t\kappa_t=0$ have been discussed. For example, under a random walk with drift model, this constraint implies $\theta$ in (\ref{eq2.2}) converges in probability to zero as $N$ goes infinite \citep{Qing}. It is also pointed out that in the presence of missing mortality rates, having a prior of $\kappa_t$ incorporated with this constraint is not applicable in that the covariance matrix of all $\kappa_t$'s except one in the full conditional is close to singular, resulting in numerical instability \citep{Johnny}. Hence, a separate step of re-scaling an MCMC sample is required in each iteration. In view of these, we adopt the constraints $\sum_x\alpha_x=0$ and $\sum_x\beta_x=1$ by \citet{Qing} into our Bayesian framework. Similar to $\sum_x\alpha_x=1$ and $\kappa_1=0$ by \citet{Jackie}, the constraints can be easily embedded into the prior specifications of $\alpha_x$'s and $\beta_x$'s. However, unlike placing $\kappa_t$-related constraints, the state equation in the Kalman filter remains simple and straightforward in expression. 
Under such a setting, $\kappa_t$ also enjoys a nice interpretation as an aggregation of log mortality rate in the $t^{th}$ year due to 
\begin{align*}
    &\sum_x\log(\mu_{x,t})\approx \sum_x\alpha_x + \sum_x\beta_x\kappa_t\\
    \Rightarrow & \kappa_t\approx \sum_x\log(\mu_{x,t}).
\end{align*}

To incorporate the constraints into the prior distributions, we start with the normal prior
\begin{align}
    \begin{bmatrix}
  \balpha\\
  \bbeta\\
  \end{bmatrix}
  \sim N \left( \begin{bmatrix}
    \bmu_{\alpha} \\
    \bmu_{\beta}\\
    \end{bmatrix},
  \; 
  \begin{bmatrix}
    \sigma_{\alpha}^2\bI_M & \mathbf {0}\\
    \mathbf {0} & \sigma_{\beta}^2\bI_M\\
    \end{bmatrix} \right), \label{eq3.4}
\end{align}
where $\balpha=(\alpha_1,\alpha_2,\dots,\alpha_{M})^T$, $\bbeta=(\beta_1,\beta_2,\dots,\beta_{M})^T$, and $\bmu_{\alpha}=(\mu_{\alpha_1}, \mu_{\alpha_2},\dots,\mu_{\alpha_M})^T$ and $ \bmu_{\beta}=(\mu_{\beta_1}, \mu_{\beta_2},\dots,\mu_{\beta_M})^T$ are pre-specified means while $\sigma_{\alpha}^2$ and $\sigma_{\beta}^2$ are the corresponding scales for variances. 
Accordingly, followed by the conditional property of a multivariate normal, (\ref{eq3.4}) subjected to the constraints can be written as 
 \begin{equation}
\label{eq3.5}
 \begin{bmatrix}
  {\balpha_{-M}}\\
  {\bbeta_{-M}}\\
  \end{bmatrix}
  \sim N\left(\bmu_p,\; \mathbold{\bSigma}_p\right),  
 \end{equation}
 where $\balpha_{-M}=(\alpha_1,\alpha_2,\dots,\alpha_{M-1})^T$, $\bbeta_{-M}=(\beta_1,\beta_2,\dots,\beta_{M-1})^T$, $\bmu_p={\bmu_1} -\bSigma_{1}\bSigma_{2}^{-1}({\bmu_2}-\mathbold{a})$, $\bSigma_p=\bSigma_3 - \mathbold{\Sigma}_{1}\bSigma_{2}^{-1}\bSigma_{1}^T$, $\bmu_1=(\bmu_{\alpha,-M}^T,\bmu_{\beta,-M}^T)^T$, 
 $\bmu_{\alpha,-M}=(\mu_{\alpha_1},\mu_{\alpha_2},\dots,\mu_{\alpha_{M-1}})^T$, $\bmu_{\beta,-M}=(\mu_{\beta_1},\mu_{\beta_2},\dots,\mu_{\beta_{M-1}})^T$, $\bmu_2=(\sum_x \mu_{\alpha_x},\sum_x \mu_{\beta_x})^T$, $\mathbold{a}=(0,1)^T$,
 $$\bSigma_{1} = \begin{bmatrix} 
    \sigma_{\alpha}^2 \mathbold{J}_{M-1} &  \mathbf{0}_{M-1}\\
      \mathbf{0}_{M-1} & \sigma_{\beta}^2 \mathbold{J}_{M-1}\\
    \end{bmatrix}, \bSigma_{2} = \begin{bmatrix}
 M\sigma_{\alpha}^2 & 0\\
 0 & M\sigma_{\beta}^2\\
 \end{bmatrix}, \mbox{ and } \bSigma_{3} = \begin{bmatrix} 
     \sigma_{\alpha}^2\bI_{M-1} & \mathbf{0}_{M-1}\mathbf{0}_{M-1}^T\\
    \mathbf{0}_{M-1}\mathbf{0}_{M-1}^T & \sigma_{\beta}^2\bI_{M-1}\\
    \end{bmatrix}.$$
Once $\balpha_{-M}$ and $\bbeta_{-M}$ are updated based on (\ref{eq3.5}) in an iteration, $\alpha_M$ and $\beta_M$ are automatically determined from $\alpha_M=-\alpha_1-\alpha_2-\dots-\alpha_{M-1}$ and $\beta_M=1-\beta_1-\beta_2-\dots-\beta_{M-1}$, respectively. 
For other parameters in the model, we propose the following priors  
\begin{align*}
    &\sigma_{\alpha}^2 \sim \text{Inv-Gamma}(a_{\sigma_{\alpha}^2},b_{\sigma_{\alpha}^2}),\\
    &\sigma_{\beta}^2 \sim \text{Inv-Gamma}(a_{\sigma_{\beta}^2},b_{\sigma_{\beta}^2}),\\
    &f(\sigma_{\varepsilon}^2)\propto\frac{1}{ \sigma_{\varepsilon}^2},\\
    &f(\theta_1) \propto 1,\\
    &\zeta \sim \text{Inv-Gamma}(a_{\zeta},b_{\zeta}),\\
    &f(\sigma_{\omega}^2)\propto \frac {1} {\sigma_{\omega}^2},
\end{align*}
where $a_{\sigma_{\alpha}^2},\, b_{\sigma_{\alpha}^2},\, a_{\sigma_{\beta}^2},\, b_{\sigma_{\beta}^2},\, a_{\zeta}$ and $b_{\zeta}$ are pre-specified hyperparameters.

%% file: Section4.tex
\section{MCMC Scheme for the Proposed Model}
In this section, we develop the posterior sampling strategies separately for the complete and incomplete data sets. The proposed algorithm is a hybrid of the Gibbs and Metropolis-Hasting samplings, and the Kalman and sequential Kalman filters. Since the full conditionals for each parameter except $\kappa_t$'s are either the same for both scenarios or can be written as functions of binary indexes $\mathbb{1}_{x,t}$'s, where $\mathbb{1}_{x,t}=0$ or 1 represents the corresponding $D_{x,t}$ is missing or observed, respectively, we first present those results in Sections 4.1 and 4.2. Followed by Sections 4.3 and 4.4, the Kalman and sequential Kalman filters are provided to update $\kappa_t$'s for the two scenarios.   


\subsection{Updating Parameters $\sigma_{\alpha}^2,$ $\sigma_{\beta}^2,$  $\theta_1,$  $z,$ $\theta_2,$  $\zeta,$ and $\sigma_{\omega}^2$}
Let the notation $``\mid."$ represent ``conditional on all other parameters and the data$"$, the full conditional distributions of $\sigma_{\alpha}^2$, $\sigma_{\beta}^2$, $\theta_1,$   $z$, $\theta_2,$  $\zeta,$ and $\sigma_{\omega}^2$ are 
\hspace{12pt} 
\begin{align*}
 &\sigma_{\alpha}^2\mid.
 \sim \text{Inv-Gamma} \left(a_{\sigma_{\alpha}^2}+\frac{M-1}{2}, b_{\sigma_{\alpha}^2}+ \frac{1}{2}\tilde{\balpha}_{-M}^T\left(\bI_{M-1}-\frac{1}{M}\bL_{M-1}\right)^{-1}\tilde{\balpha}_{-M}\right),\\
 & \sigma_{\beta}^2\mid.
 \sim \text{Inv-Gamma} \left(a_{\sigma_{\beta}^2}+\frac{M-1}{2},b_{\sigma_{\beta}^2}+ \frac{1}{2}\tilde{\bbeta}_{-M}^T\left(\bI_{M-1}-\frac{1}{M}\bL_{M-1}\right)^{-1}\tilde{\bbeta}_{-M}\right),\\
 & \theta_1\mid.\sim N \left(\frac{\kappa_{N}-\kappa_{0}-\theta_2\sum_{t}t}{N}, \; \frac{\sigma_{\omega}^2}{N}\right),\\
 & z\mid . \sim \text{Bernoulli}(\tilde{p}), \\
 & \theta_2\mid.\sim \begin{cases}
  N\left(\frac{\zeta\sum_t{t(\kappa_t-\kappa_{t-1}-\theta_1})}{\sigma_{\omega}^2+\zeta\sum_t{t^2}}, \; \frac{\zeta\sigma_{\omega}^2}{\sigma_{\omega}^2+\zeta\sum_t{t^2}}\right) & \quad \text{if } z = 1,\\
  \pi & \quad \text{if } z = 0,
 \end{cases}\\
 & \zeta\mid.\sim\begin{cases}
     \text{Inv-Gamma}\left(a_{\zeta} + \frac{1}{2},\;b_{\zeta}+\frac{\theta_2^2}{2}\right) & \quad \text{if } z = 1,\\
     \text{Inv-Gamma}\left(a_{\zeta},\;b_{\zeta}\right) & \quad \text{if } z = 0,
     \end{cases}\\
 & \sigma_{\omega}^2\mid . \sim \text{Inv-Gamma} \left(\frac{N}{2},\; \frac{\sum_{t}(\kappa_t - \kappa_{t-1} - \theta_1-\theta_2t)^2}{2}\right),
\end{align*}
where $\tilde{\balpha}_{-M}=\balpha_{-M}-\bmu_{\alpha,-M}$, $\tilde{\bbeta}_{-M}=\bbeta_{-M}-\bmu_{\beta,-M}$, $\bL_{M-1}=\bJ_{M-1}\bJ_{M-1}^T$.

\subsection{Updating Parameters $\log(\mu_{x,t}),$ $\alpha_x,$ $\beta_x$ and $\sigma_{\varepsilon}^2$}
Essentially, the M-H procedure for $\log(\mu_{x,t})$ is the same as the steps in Section 3.1 except that the modified full conditional distribution is needed to accommodate the situation when $D_{x,t}$ is missing. Specifically, we incorporate $\mathbb{1}_{x,t}$ into (\ref{logmu})
\begin{align*}
\begin{split}
    \pi\left(\log (\mu_{x,t})\mid .\right)
     \propto \mu_{x,t}^{D_{x,t}}\;\exp\left[-E_{x,t} \mu_{x,t}-\frac{1}{2\sigma_{\varepsilon}^2} \left(\log(\mu_{x,t})-\alpha_x-\beta_x\kappa_t\right)^2\right]\mathbb{1}_{x,t}.
     \end{split}
\end{align*}
It is clear when $\mathbb{1}_{x,t}=0$, $\log(\mu_{x,t})$ can not be updated due to unavailability of $D_{x,t}$. As $\mathbb{1}_{x,t}=1$, we follow the trial and error method in Section 3.1 to find the ideal value of $\sigma_{x,t}^2$ for the proposed density.


Let $\by_{-M,t}^*=(\log(\mu_{1,t})\mathbb{1}_{1,t}, \log(\mu_{2,t})\mathbb{1}_{2,t},\dots,\log(\mu_{M-1,t})\mathbb{1}_{M-1,t})^T$, and following the prior setting in (\ref{eq3.5}), $\alpha_x$ and $\beta_x$ are updated via 
 \begin{align*}
 \begin{cases}
    &\balpha_{-M},\bbeta_{-M}\mid . \sim N (\tilde{\bmu}_p, \tilde{\bSigma}_p),\\
    &\alpha_M=-\alpha_1-\alpha_2-\dots-\alpha_{M-1},\\
    &\beta_M=1-\beta_1-\beta_2-\dots-\beta_{M-1}, 
\end{cases}
\end{align*}
 where $\tilde{\bmu}_p=\tilde{\bSigma}_p(\bmu_d + \bmu_p^T \bSigma_p^{-1})$, $\tilde{\bSigma}_p=(\bSigma_d+\bSigma_p^{-1})^{-1}$, $\bmu_d=(\bmu_{d_1}^T,\bmu_{d_2}^T)^T$, $\bSigma_d=
     \begin{bmatrix}
    \mathbf {A} & \mathbf {B}\\
    \mathbf B & \mathbf C
    \end{bmatrix}$, 
     \begin{align*}
     &\bmu_{d_1}=\frac{1}{\sigma_{\varepsilon}^2}\left[\sum_t \by_{-M,t}^*+\sum_t\left(\kappa_t-\log(\mu_{M,t})\right)\mathbb{1}_{M,t}\times \bJ_{M-1}\right],\\
     &\bmu_{d_2}=\frac{1}{\sigma_{\varepsilon}^2}\left[\sum_t \kappa_t\times \by_{-M,t}^*+\sum_t\kappa_t \left(\kappa_t-\log(\mu_{M,t})\right)\mathbb{1}_{M,t}\times \bJ_{M-1}\right],\\
     &\mathbf{A} = \begin{bmatrix}
\sum_t{\mathbb{1}_{1,t}} &  0  & \ldots & 0\\
0  &  \sum_t{\mathbb{1}_{2,t}} & \ldots & 0\\
\vdots & \vdots & \ddots & \vdots\\
0  &   0       &\ldots & \sum_t{\mathbb{1}_{M-1,t}}
\end{bmatrix}+\left(\sum_t{\mathbb{1}_{M,t}}\right)\times\bJ_{M-1}\bJ_{M-1}^T,\\
&\mathbf{B} =
 \begin{bmatrix}
\sum_t{\kappa_t\mathbb{1}_{1,t}} &  0  & \ldots & 0\\
0 &  \sum_t{\kappa_t\mathbb{1}_{2,t}}  & \ldots & 0\\
\vdots & \vdots & \ddots & \vdots\\
0  &   0       &\ldots & \sum_t{\kappa_t\mathbb{1}_{M-1,t}}
\end{bmatrix}+\left(\sum_t{\kappa_t\mathbb{1}_{M,t}}\right)\times\bJ_{M-1}\bJ_{M-1}^T,
\end{align*}
and
\begin{align*}
&\mathbf{C} = \begin{bmatrix}
\sum_t{\kappa_t^2\mathbb{1}_{1,t}} &  0  & \ldots & 0\\
0  &  \sum_t{\kappa_t^2\mathbb{1}_{2,t}} & \ldots & 0\\
\vdots & \vdots & \ddots & \vdots\\
0  &   0       &\ldots & \sum_t{\kappa_t^2\mathbb{1}_{M-1,t}}
\end{bmatrix}+\left(\sum_t{\kappa_t^2\mathbb{1}_{M,t}}\right)\times\bJ_{M-1}\bJ_{M-1}^T.
     \end{align*}
As for $\sigma_{\varepsilon}^2$, its full conditional is conjugate and given by
 \begin{align*}
\sigma_{\varepsilon}^2\mid. \sim \text{Inv-Gamma} \left(\frac{\sum_x\sum_t{\mathbb{1}_{x,t}}}{2},\; \frac{\sum_x\sum_t\left(\log(\mu_{x,t}) - \alpha_x - \beta_x\kappa_t\right)^2\mathbb{1}_{x,t}}{2}\right).
\end{align*}

 \subsection{Kalman filter for $\mathbold{\kappa}$}
For the complete data set, we can rewrite (\ref{eq3.2}) and (\ref{eq3.3}) as the 
state space form (or known as dynamical linear model). The so-called observation and state equations are separately given by
\begin{align}
    \by_t = \bF_t\bm_{t}+ \bv_t,\qquad \bv_t \sim N(\bzero_M,\,\bV_t),&\\
        \bm_t = \bG_t\bm_{t-1}+ \bw_t,\quad \bw_t \sim N(\bzero_3,\,\bW_t),
\end{align}
where $\by_t = (\log(\mu_{1,t}), \log(\mu_{2,t}), \dots, \log(\mu_{M,t}))^T$, $\bm_t=(1, \kappa_t, t+1)^T$, $\bv_t = (\varepsilon_{1,t}, \varepsilon_{2,t}, \dots, \varepsilon_{M,t})^T$, $\bw_t=(0,\omega_t, 0)^T$, $\bF_t=\begin{bmatrix}\balpha & \bbeta & \bzero_M\end{bmatrix}$, $\bV_t=\sigma_{\varepsilon}^2\times \bI_M$, 
 \begin{flalign*}
 \bG_t = \begin{bmatrix}
 1 & 0 & 0 \\
 \theta_1 & 1 & \theta_2\\t+1 & 0 &0
 \end{bmatrix}, \mbox{ and }
 \bW_t = \begin{bmatrix}
 0 & 0 & 0\\ 0 & \sigma_{\omega}^2 & 0\\0 & 0 & 0
 \end{bmatrix}.
 \end{flalign*}
In the Gibbs sampler, $\by_t$ is readily used to improve the full conditional of $\kappa_t$ in the following manner. Let $D_{1\mapsto t}=\{\by_1, \by_2, \dots, \by_t \}$ denote the data containing all up-to-now information, 
 and define 
$``\mid._{(1\mapsto t)}"=``\mid \balpha, \bbeta, \theta_1,\theta_2, \sigma^2_{\varepsilon}, \sigma^2_{\omega}, \kappa_1,\kappa_2,\dots, \kappa_{t-1}, D_{1\mapsto t}"$. The filtering process utilizes $D_{1\mapsto t}$ to recursively update $\kappa_t$ from $t=1$ to $t=N$. Specifically, we have $\bm_t\mid ._{(1\mapsto t)}\sim N\left(\E(\bm_t\mid._{(1\mapsto t)} ),\var(\bm_t\mid._{(1\mapsto t)} )\right)$ with 
\begin{align}
\begin{split}
   &\E(\bm_t\mid._{(1\mapsto t)} )=\bG_t\E(\bm_{t-1}\mid._{(1\mapsto t-1)} ) + \bK_t\be_t,\\
&\var(\bm_t\mid._{(1\mapsto t)} )=\bR_t-\bK_{t}\mathbf{F}_t\mathbf{R}_{t}, \label{filter}
\end{split}
\end{align}
where $\bK_t = \bR_t\bF_t^T(\bV_t + \bF_t\bR_t\bF_t^T)^{-1}$, $\be_t = \by_t-\bF_{t}\bG_t\E(\bm_{t-1}\mid._{(1\mapsto t-1)} )$, $\bR_t = \bG_{t}\var(\bm_{t-1}\mid._{(1\mapsto t-1)} )\bG_{t}^T+\bW_t$. Since $\be_t$ measures the difference between true $\by_t$ and its expectation based on all other parameters and data up to time $t-1$ except $\kappa_{t-1}$, it provides a correction transformed by $\bK_t$ on $\E(\bm_t\mid._{(1\mapsto t)} )$. 
Without this correction, the conditional expectation is merely derived from the state equation. Similarly, $\bF_t\bR_t$ plays the same role on $\var(\bm_t\mid._{(1\mapsto t)} )$ via $\bK_t$. Also, note that $\E(\bm_0\mid\balpha, \bbeta, \theta_1,\theta_2, \sigma^2_{\varepsilon}, \sigma^2_{\omega} )$ and $\var(\bm_0\mid\balpha, \bbeta, \theta_1,\theta_2, \sigma^2_{\varepsilon}, \sigma^2_{\omega} )$ are estimated by  the LC model using SVD approach. 

Once $\kappa_N$ is retained from the forward algorithm above, it 
is reversely integrated to the conditional mean and covariance of the previous term in the smoothing process. 
Proceeding in this fashion, we obtain $\kappa_t$ sequentially from $t=N-1$ to $t=1$ via $\bm_t\mid ._{(1\mapsto t)},\kappa_{t+1}\sim N\left(\E(\bm_t\mid._{(1\mapsto t)},\kappa_{t+1} ), \var(\bm_t\mid._{(1\mapsto t)},\kappa_{t+1} )\right)$ with  
\begin{align}
    &\E(\bm_t\mid._{(1\mapsto t)},\kappa_{t+1} )\nonumber\\
    =&\E(\bm_t\mid._{(1\mapsto t)} ) + \var(\bm_t\mid._{(1\mapsto t)} )\bG_{t+1}^T\bR_{t+1}^{-1}\left[\bm_{t+1} -\bG_{t+1}\E(\bm_t\mid._{(1\mapsto t)} )\right]\label{eq4.3}
\end{align}
and 
\begin{align}
    &\var(\bm_t\mid._{(1\mapsto t)},\kappa_{t+1} )\nonumber\\
    =&\var(\bm_t\mid._{(1\mapsto t)} ) - \var(\bm_t\mid._{(1\mapsto t)} )\bG_{t+1}^T\bR_{t+1}^{-1}\bG_{t+1}\var(\bm_t\mid._{(1\mapsto t)} ).\label{eq4.4}
\end{align}
Now, $\kappa_{N}$ from the filtering process and $\kappa_{N-1},\kappa_{N-2},\dots,\kappa_1$ from the smoothing process jointly form one iteration in the MCMC sample.



\subsection{Sequential Kalman filter for $\mathbold{\kappa}$}

\hspace{12pt} In the presence of missing data, due to unavailability of some $\log{\mu_{x,t}}$'s, the aforementioned filtering process that aims to update all information of $\by_t$ at once to the conditional mean and variance of $\kappa_t$ is hindered from used. To this end, we modify (\ref{filter}) as the sequential Kalman filter \citep{Durbin} to allow $\log{\mu_{1,t}},\log{\mu_{2,t}},\dots,\log{\mu_{M,t}}$ sequentially 
formulating the conditional structure of $\kappa_t$ in the forward algorithm. First, let $\mathcal{D}_{1\mapsto t_x}=\{\by_1, \by_2, \dots, \by_{t-1}, \log(\mu_{1,t}), \log(\mu_{2,t}),\dots,\log(\mu_{x,t}) \}$ 
and $``\mid._{(1\mapsto t_x)}"=``\mid \balpha, \bbeta, \theta_1,\theta_2, \sigma^2_{\varepsilon}, \sigma^2_{\omega}, \kappa_1,\kappa_2,\dots, \kappa_{t-1}, \mathcal{D}_{1\mapsto t_x}"$. Also, 
rewrite (\ref{eq3.2}) and  (\ref{eq3.3}) as
\begin{align*}
    &\log(\mu_{x,t})=\alpha_x+\beta_x\kappa_{x,t}+\varepsilon_{x,t},
\end{align*}
and
\begin{align*}
    &\kappa_{x,t}=
    \begin{cases}
    \theta_1 +\theta_2t + \kappa_{M,t-1} + \omega_t \quad & \text{if } x=1,\\
    \kappa_{x-1,t} \quad & \text{if }
    1 < x \leq M,
    \end{cases}
\end{align*}
where $\kappa_{x,t}$ remains constant across different age groups at time $t$ and can be viewed as a hidden state affecting $\log(\mu_{x,t})$. In return, this observed $\log(\mu_{x,t})$ can be used to determine the conditional structure of $\kappa_{x,t}$, i.e. $\E(\kappa_{x,t}\mid._{(1\mapsto t_x)})$ and $\var(\kappa_{x,t}\mid._{(1\mapsto t_x)})$.
Accordingly, we develop the following recursive equations for our proposed model: when $x=1$ and $\sum_x\mathbb{1}_{x,t}<M$, 
\begin{align}
    \begin{split}
        &\E(\kappa_{x,t}\mid._{(1\mapsto t_x)}) = \theta_1 +\theta_2t + \E(\kappa_{M,t-1}\mid._{(1\mapsto (t-1)_M)}) + k_{x,t} e_{x,t}\mathbb{1}_{x,t},\\
        &\var(\kappa_{x,t}\mid._{(1\mapsto t_x)}) = (1- k_{x,t}\beta_{x}\mathbb{1}_{x,t})\left[\var(\kappa_{M,t-1}\mid._{(1\mapsto (t-1)_M)}) + \sigma_{\omega}^2\right]; \label{skf1}
    \end{split}
\end{align}
when $1 < x \leq M$ and $\sum_x\mathbb{1}_{x,t}<M$, 
\begin{align}
    \begin{split}
        &\E(\kappa_{x,t}\mid._{(1\mapsto t_x)}) = \E(\kappa_{x-1,t}\mid._{(1\mapsto t_{x-1})}) + k_{x,t} e_{x,t}\mathbb{1}_{x,t},\\
     &\var(\kappa_{x,t}\mid._{(1\mapsto t_x)}) = (1- k_{x,t}\beta_{x}\mathbb{1}_{x,t})\var(\kappa_{x-1,t}\mid._{(1\mapsto t_{x-1})}),  \label{skf2}
    \end{split}
\end{align}
where 
\begin{align*}
    \begin{split}
    &e_{x,t} =     \begin{cases}
        \log(\mu_{x,t}) - \alpha_x -\beta_x\left[\theta_1+\theta_2t + \E(\kappa_{M,t-1}\mid._{(1\mapsto (t-1)_M)})\right] \quad & \text{if } x=1 \mbox{ and } \mathbb{1}_{x,t}=1,\\
        \log(\mu_{x,t}) - \alpha_x -\beta_x \E(\kappa_{x-1,t}\mid._{(1\mapsto t_{x-1})}) \quad & \text{if } 1 < x \leq M \mbox{ and } \mathbb{1}_{x,t}=1,\\
        0 \quad & \text{if } \mathbb{1}_{x,t}=0,\\
        \end{cases}\\
    &k_{x,t} = \begin{cases}
     \frac{\beta_{1}}{\beta_{1}^2+\sigma_{\varepsilon}^2\left[\var(\kappa_{M,t-1}\mid._{(1\mapsto (t-1)_M)})+\sigma_{\omega}^2\right]^{-1}} \quad & \text{if } x=1 \mbox{ and } \mathbb{1}_{x,t}=1,\\
     \frac{\beta_{x}}{\beta_{x}^2+\sigma_{\varepsilon}^2\left[\var(\kappa_{x-1,t}\mid._{(1\mapsto t_{x-1})})\right]^{-1}} \quad & \text{if } 1 < x \leq M \mbox{ and } \mathbb{1}_{x,t}=1,\\
     0 \quad & \text{if } \mathbb{1}_{x,t}=0.
     \end{cases} 
    \end{split}
\end{align*}
As for any year with $\sum_x\mathbb{1}_{x,t}=M$, the sequential update procedure in (\ref{skf1}) and (\ref{skf2}) is not necessary. Instead, (\ref{filter}) can be implemented directly to obtain $\E(\kappa_{M,t}\mid._{(1\mapsto t_M)})$ and $\var(\kappa_{M,t}\mid._{(1\mapsto t_M)})$. Note that switching back and forth between filtering and sequential filtering processes depends on the missing status in each year and is permitted because $``\mid._{(1\mapsto t)}"= ``\mid._{(1\mapsto t_M)}"$, $\E(\kappa_{t}\mid._{(1\mapsto t)})=\E(\kappa_{M,t}\mid._{(1\mapsto t_M)})$, $\var(\kappa_{t}\mid._{(1\mapsto t)})=\var(\kappa_{M,t}\mid._{(1\mapsto t_M)})$, and conditional means and variances of $\kappa_t$ and $\bm_t=(1,\kappa_t,t+1)^T$ can be retrieved from each other. 
Once $\E(\kappa_{M,N}\mid._{(1\mapsto N_M)})$ and $\var(\kappa_{M,N}\mid._{(1\mapsto N_M)})$ are obtained, an MCMC iteration of $\kappa_M$ is generated and initiates the smoothing process in (\ref{eq4.3}) and (\ref{eq4.4}) to finalize the updates of $\bkappa$. It also worth mentioning that regarding to the initial values of $\E(\kappa_0\mid\balpha, \bbeta, \theta_1,\theta_2, \sigma^2_{\varepsilon}, \sigma^2_{\omega} )$ and $\var(\kappa_0\mid\balpha, \bbeta, \theta_1,\theta_2, \sigma^2_{\varepsilon}, \sigma^2_{\omega} )$, \citet{Claudia} suggests using the empirical results from past relevant works while 
\citet{Johnny} adopts diffuse initial values. Here, we propose to 
first restore the missing values, and on this base, apply the LC model with SVD 
to retain the starting points.

%% file: Section5.tex
\section{Data Application}
\subsection{Data}
\begin{figure}[H]
 \centering
\begin{minipage}{1\textwidth}
\begin{tikzpicture}
  \node (img)  {\includegraphics[scale=0.5]{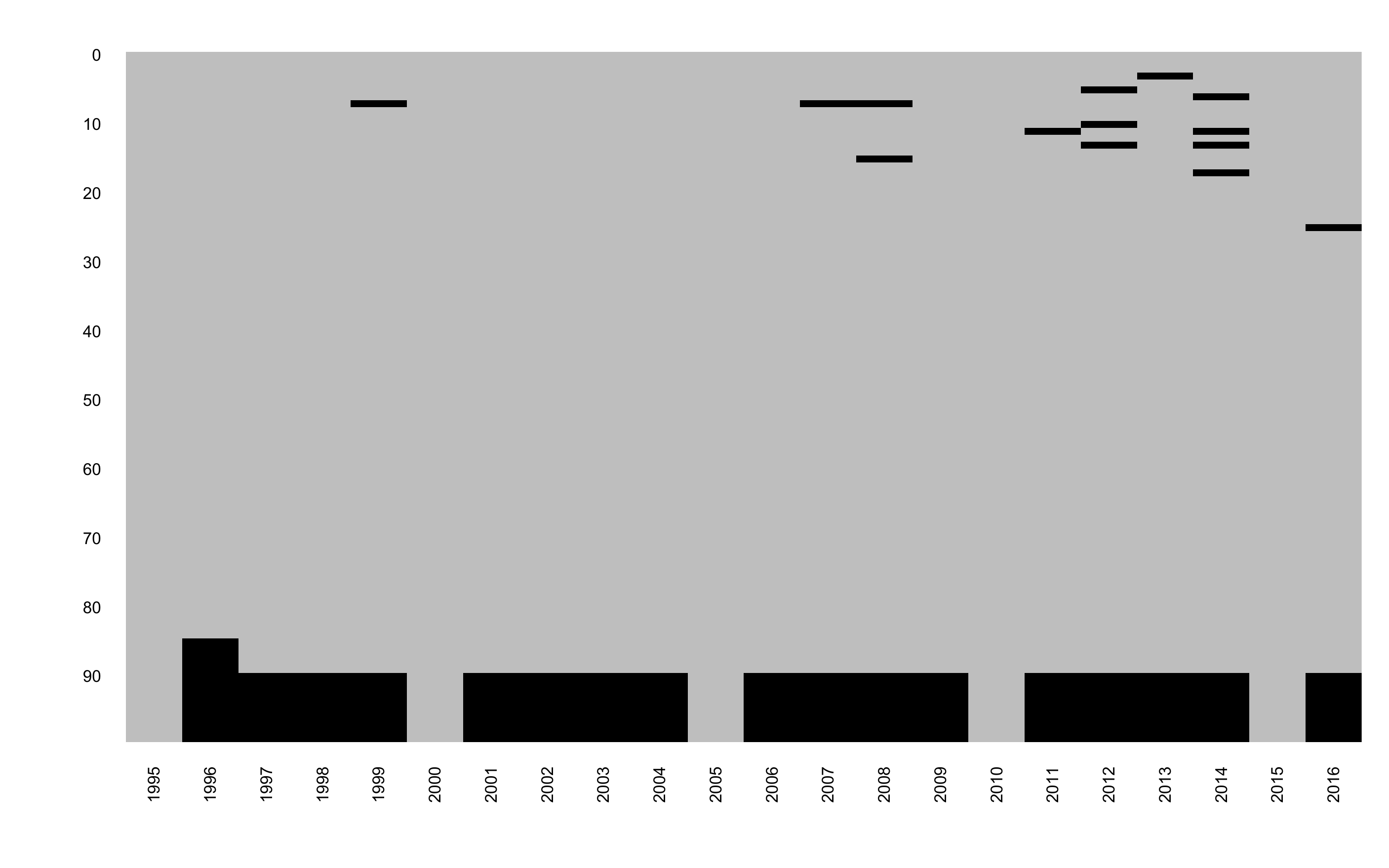}};
  \node[below=of img, node distance=0cm,  xshift=1cm, yshift=1.6cm,font=\small] {Year};
  \node[left=of img, node distance=0cm,  anchor=center,xshift=1.2cm,font=\small, rotate =90] {Age};
 \end{tikzpicture}
\end{minipage}
\caption{\label{fig:0} Availability of 2,200 age-year exposures and death counts of Chinese male, where a grey or black cell stands for the observed or missing rate of the corresponding age and year, respectively.}
\end{figure}

The data used for illustration is 1996-2017 Chinese male mortality data from China {\it Population and Employment Statistics Yearbooks} published by the \citet{China}. Due to the relatively large population size and budget limitation, nationwide censuses were only carried out in the years 2000 and 2010 within this 22-years window while $1\%$ of national population were surveyed via the multistage cluster sampling scheme for the years 1995, 2005 and 2015. For the rest of years, surveys were only performed as a small scale in $0.1\%$ of the national population. This collection procedure 
partially explains the missing patterns of the data set. As shown in \autoref{fig:0}, which depicts the availability of death counts for 2,200 age-year (100 age and 22 year groups) cells of the nationwide male, all missing values occur at the years with small scale surveys, and the missing data problem becomes severe in the senility and adolescent likely due to the minority elders in nature or lower accessibility to teenagers. 
For such reasons, we believe that the mortality rates are missing by chance, and the 
results based on the incomplete data should be similar to the ones based on imputed complete data. We present both estimations and predictions in Section 5.3 to backup this speculation.     

 \captionsetup[figure]{labelfont={bf},labelformat={default},labelsep=period,name={Fig.}}
 
\subsection{Initial Values and Computational Specifications} 
As indicated by \citet{Carpenter}, the rate of convergence of a Gibbs sampling is sensitive to the initial values when the data set contains missing observations. To avoid this potential issue, we develop a new searching procedure to train the starting points for the incomplete data analysis.
First, we use linear interpolations to impute all missing $D_{x,t}$'s and $E_{x,t}$'s. 
Supposed $t^{(0)}$ and $t^{(1)}$ denote two most adjacent years such that $t\in (t^{(0)}, t^{(1)})$, and $D_{x,t^{(0)}}$ and $D_{x,t^{(1)}}$ are available, the imputed death count is then given by 
\begin{align}
\label{5.1}
D_{x,t} = \left(\frac{t-t^{(0)}}{t^{(1)}-t^{(0)}}\right)D_{x,t^{(1)}} +\left(\frac{t^{(1)}-t}{t^{(1)}-t^{(0)}}\right)D_{x,t^{(0)}}.
 \end{align}
The similar approach is applied to the missing exposures. 
Next, based on the complete data set, the LC model with SVD is implemented to improve the previous imputations, where 
(\ref{5.1}) is now  replaced by $D_{x,t} = \hat{\mu}_{x,t}E_{x,t}$, 
and $\hat{\mu}_{x,t}$ is retrieved from (\ref{eq3.2}) and the SVD parameter estimates. 
In the final step, we use these SVD estimates to initiate the MCMC sampling of the proposed model on the new complete data set. We then choose the posterior mean of each parameter as its initial value. 

For the prior specifications, we set  $a_{\sigma_{\alpha}^2} = b_{\sigma_{\alpha}^2} = a_{\sigma_{\beta}^2} = b_{\sigma_{\beta}^2} = 0.01$ for the age-related hyper-parameters, and $a_{\zeta}$ = $b_{\zeta} = 0.1$ and $p=0.5$ in the dirac spike. The proposed values of $\sigma^2_{x,t}$'s and the corresponding acceptance rates in the last cycle of a pre-burn-in sampling are also summarized in Figure 2. It is clear that all proposed densities for $\log(\mu_{x,t})$'s end up with reasonable rates.  
 
 \begin{figure}[H]
 \centering
\begin{minipage}{1\textwidth}
\begin{tikzpicture}
  \node (img)  {\includegraphics[scale=0.5]{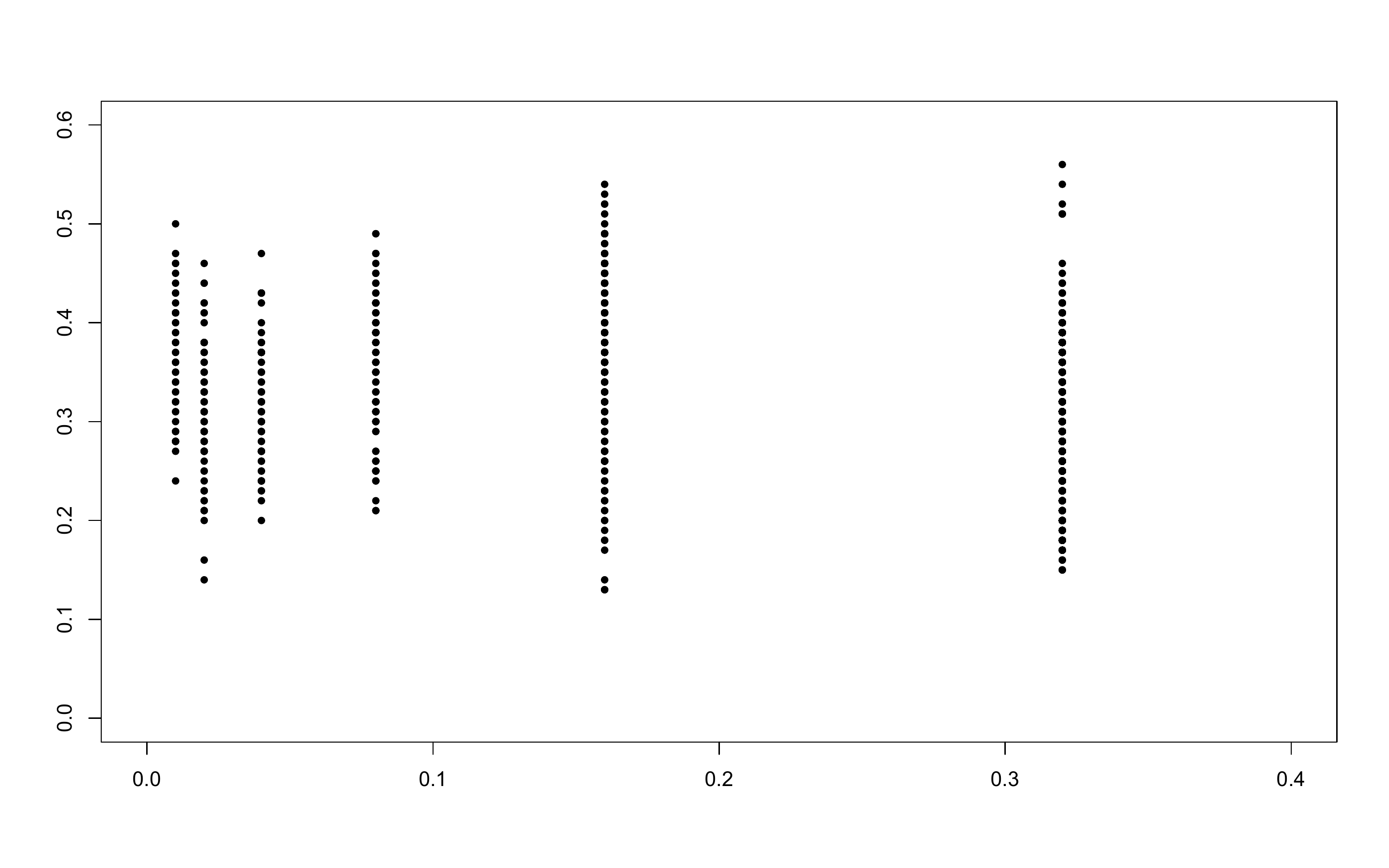}};
    \node[below=of img, node distance=0cm,  xshift=0cm, yshift=1.6cm,font=\small] {$\sigma_{x,t}^2$};
  \node[left=of img, node distance=0cm,  anchor=center,xshift=1.2cm,font=\small, rotate =90] {Acceptance Rate};
 \end{tikzpicture}
\end{minipage}
\caption{\label{fig:x} The proposed variances and their corresponding acceptance rates in the last 100-iteration cycle of the pre-burn-in sampling.}
\end{figure}

\subsection{Results} 
To construct the mortality projections of Chinese male, we begin with the steps in Section 5.2 to explore the initial values, and then fit the proposed model to the original mortality data without imputations. Based on the generated MCMC sample of 2,000 iterations after 100 burn-ins, we first examine the MCMC convergence in Figure 3, where the trace plots (left column) of five selected parameters $\alpha_{51}$, $\beta_{51}$, $\kappa_5$, $\sigma_{\varepsilon}^2$, and $\sigma_\omega^2$ are presented along with the results (right column) initiated by the SVD estimates. 
We can see that the MCMC sample following the suggested initial-value procedure has overall better convergence in the sense that each MCMC chain only requires around 50 iterations to stabilize. By contrast, the chains without the proper initial-value proposals may require more iterations as shown in the three bottom right plots.

\begin{figure}[H]
\begin{tikzpicture}
  \node (img1)  {\includegraphics[scale=0.23]{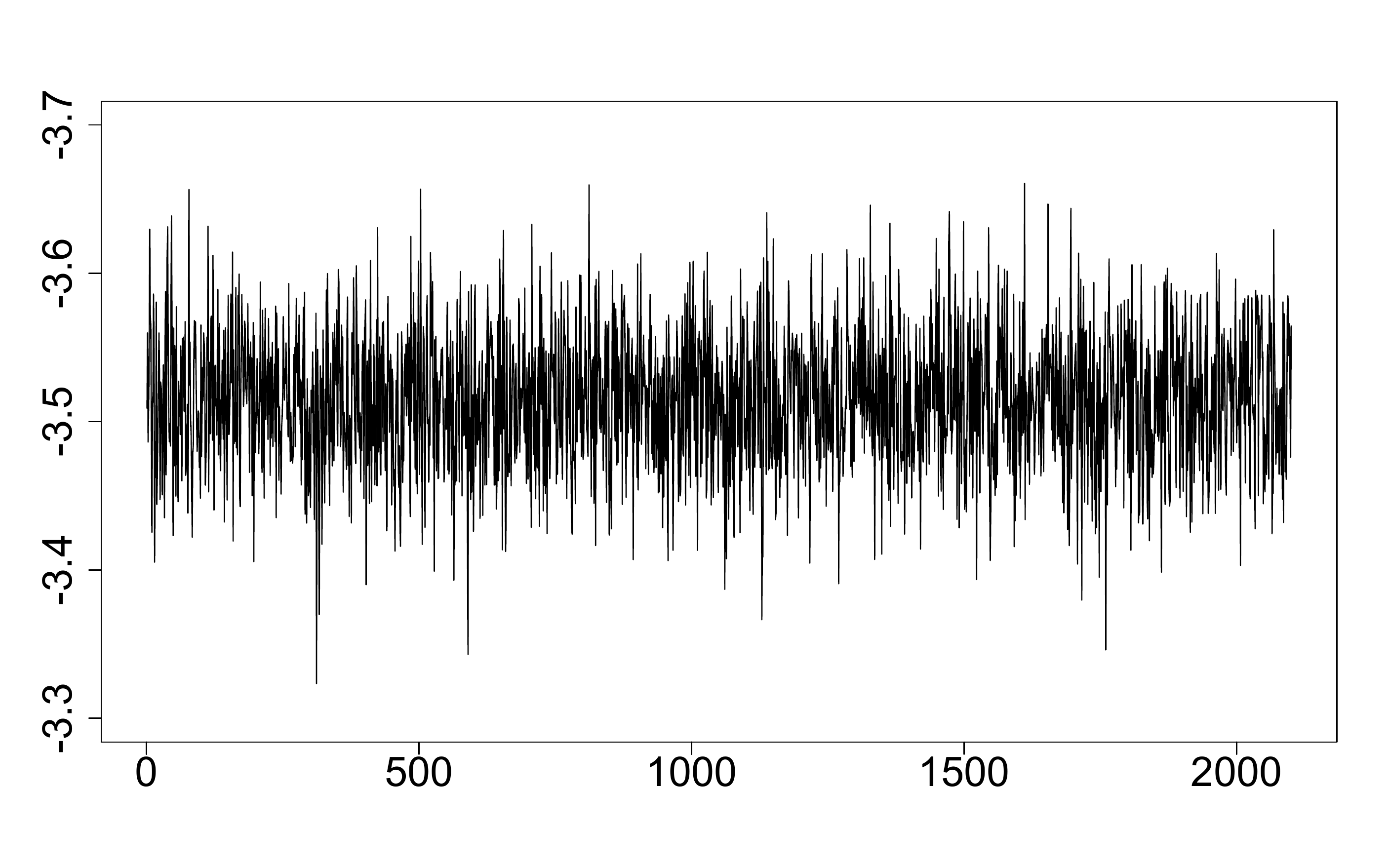}};
    \node[below=of img1, yshift=1cm] (img2)  {\includegraphics[scale=0.23]{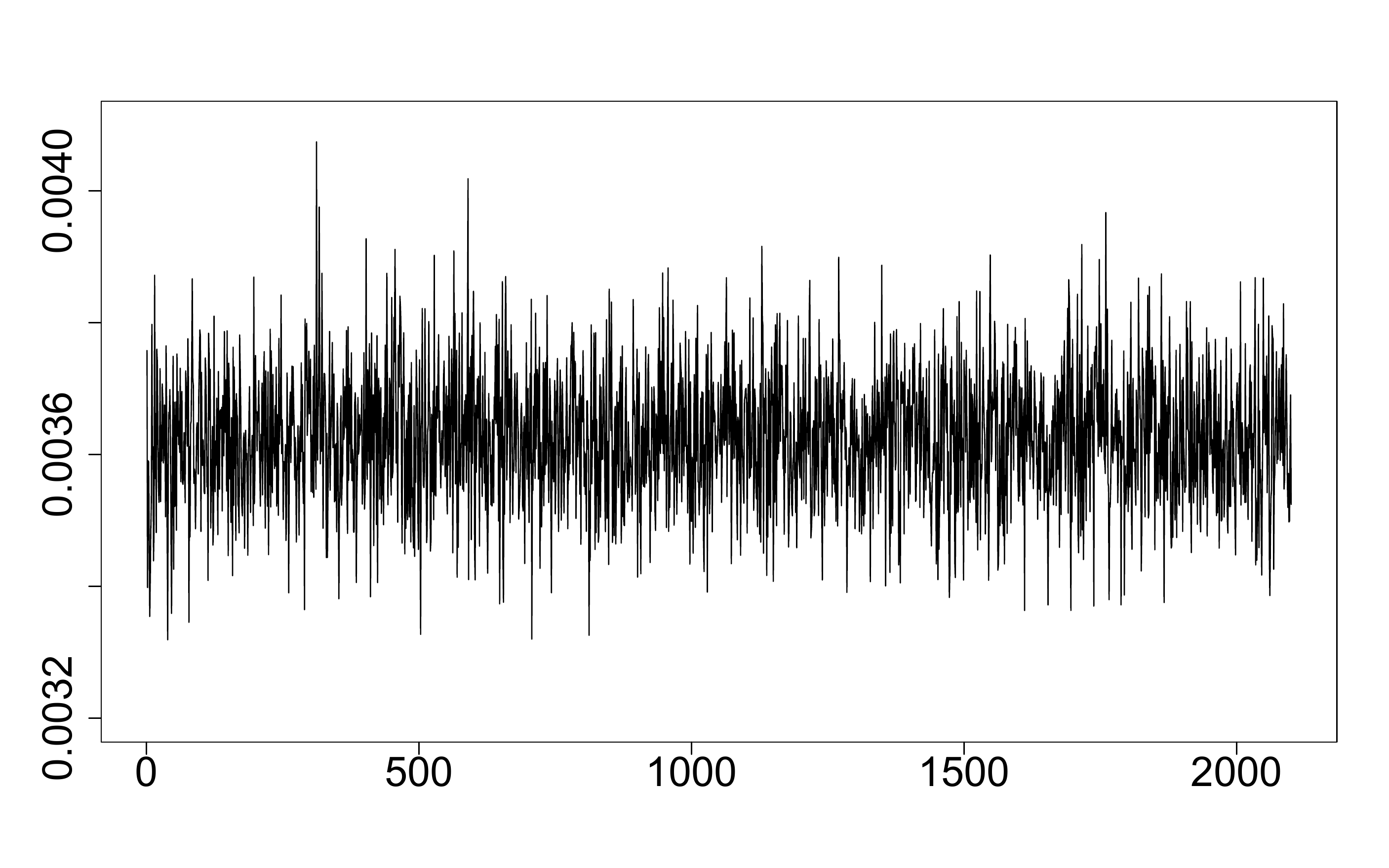}};
    \node[below=of img2, yshift=1cm] (img3)  {\includegraphics[scale=0.23]{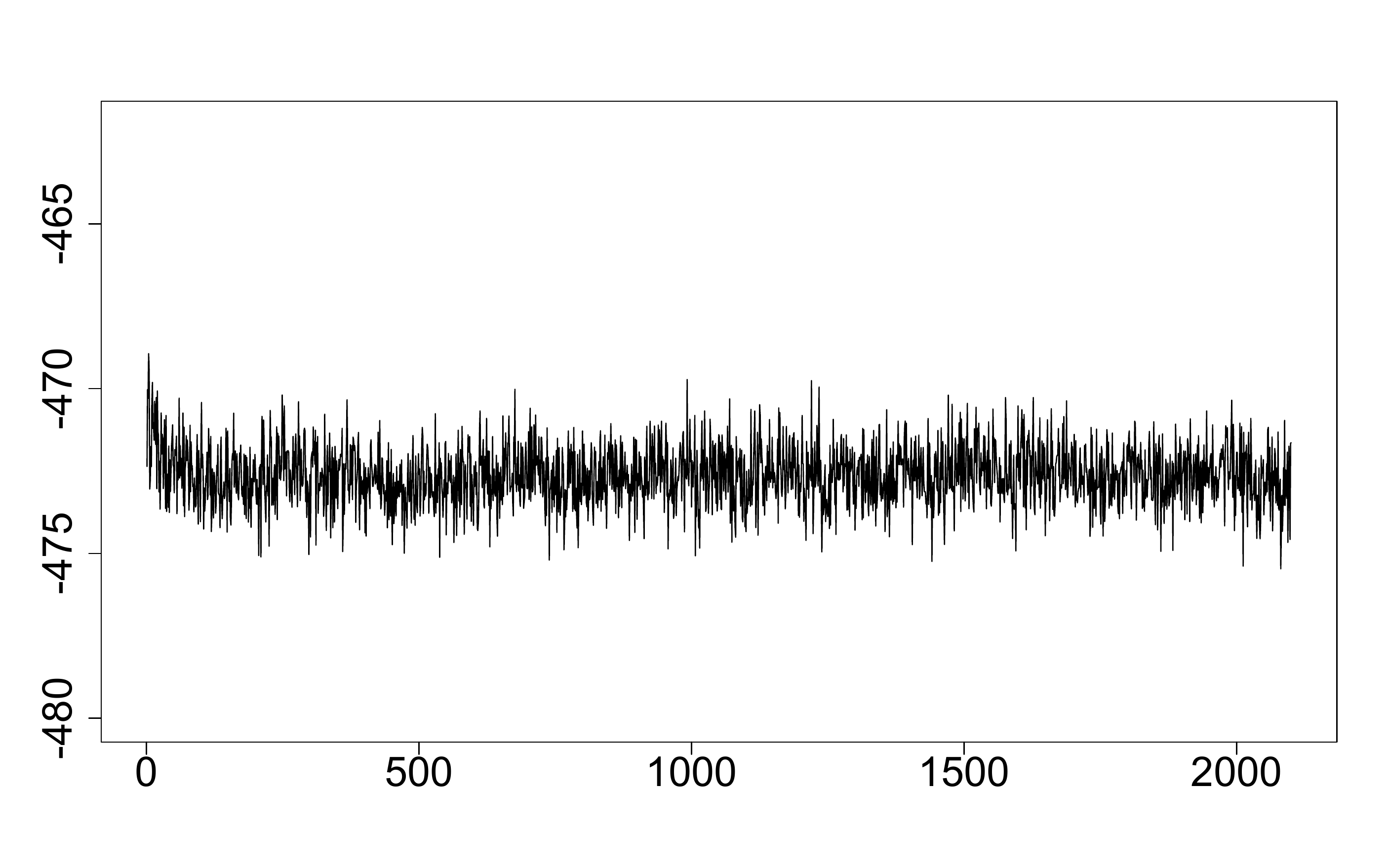}};
    \node[below=of img3, yshift=1cm] (img4)  {\includegraphics[scale=0.23]{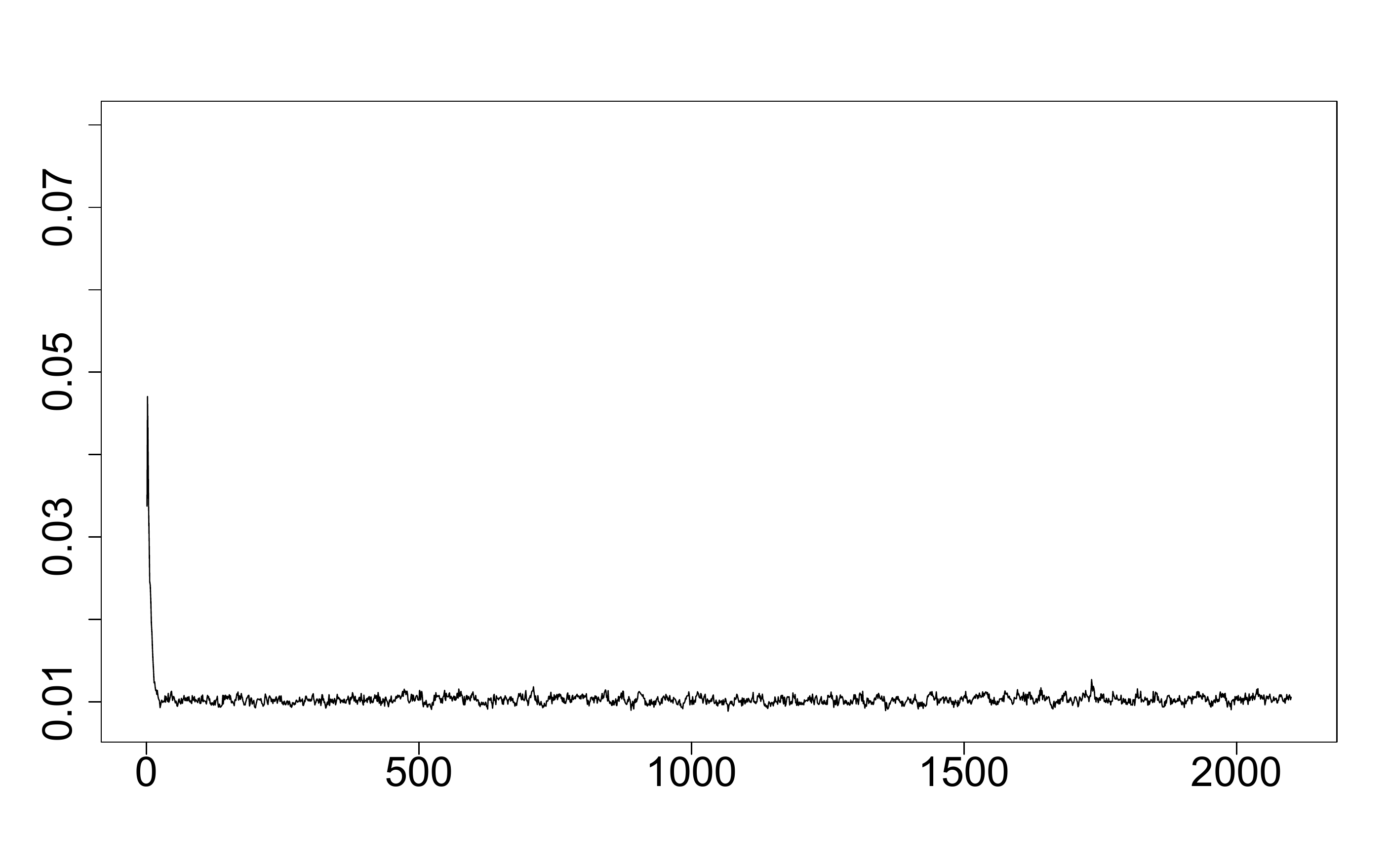}};  
    \node[below=of img4, yshift=1cm] (img5)  {\includegraphics[scale=0.23]{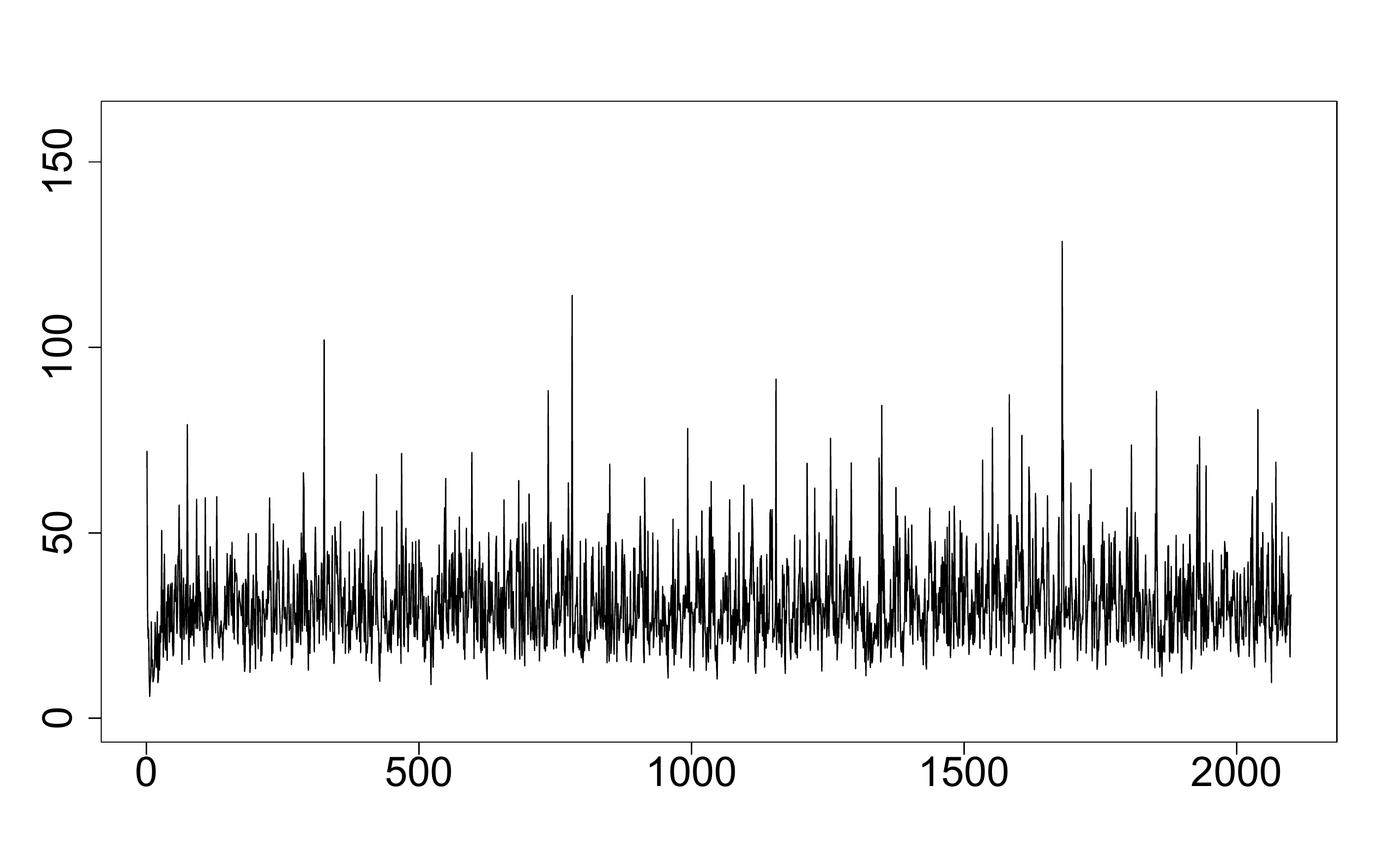}}; 
    
    \node[right=of img1, xshift=-1cm] (img6)  {\includegraphics[scale=0.23]{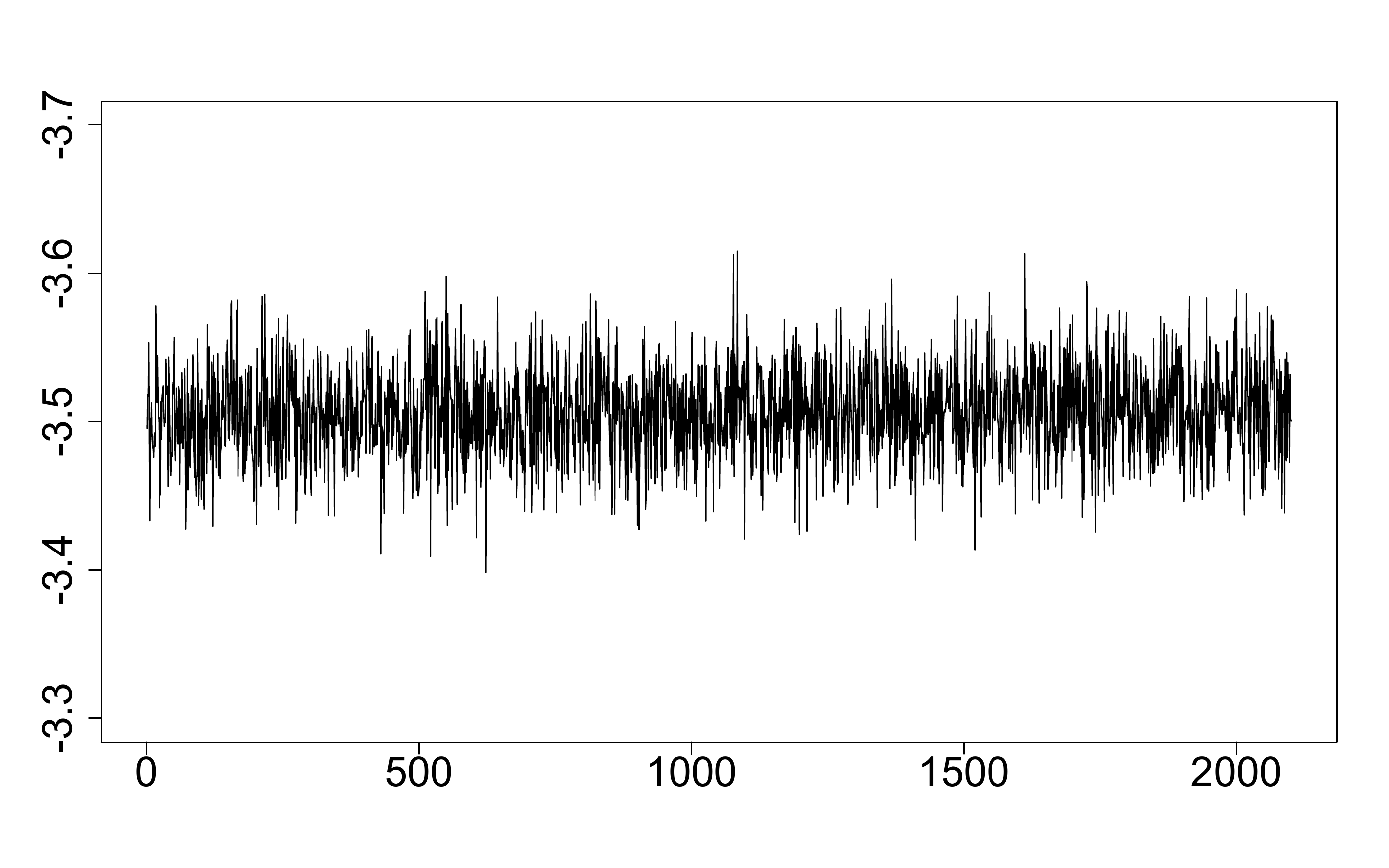}};
    \node[below=of img6,yshift=1cm] (img7)  {\includegraphics[scale=0.23]{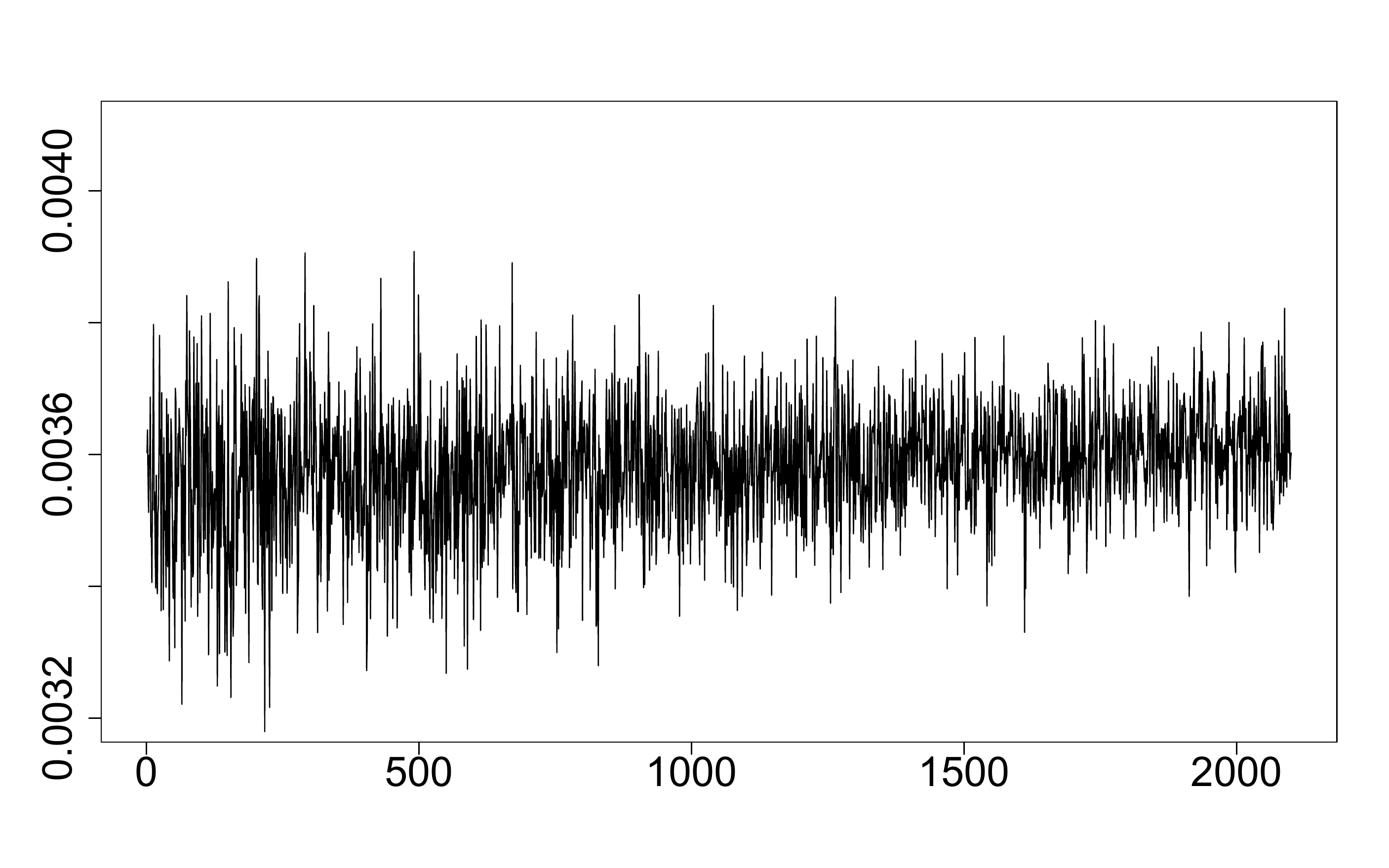}};
    \node[below=of img7,yshift=1cm] (img8)  {\includegraphics[scale=0.23]{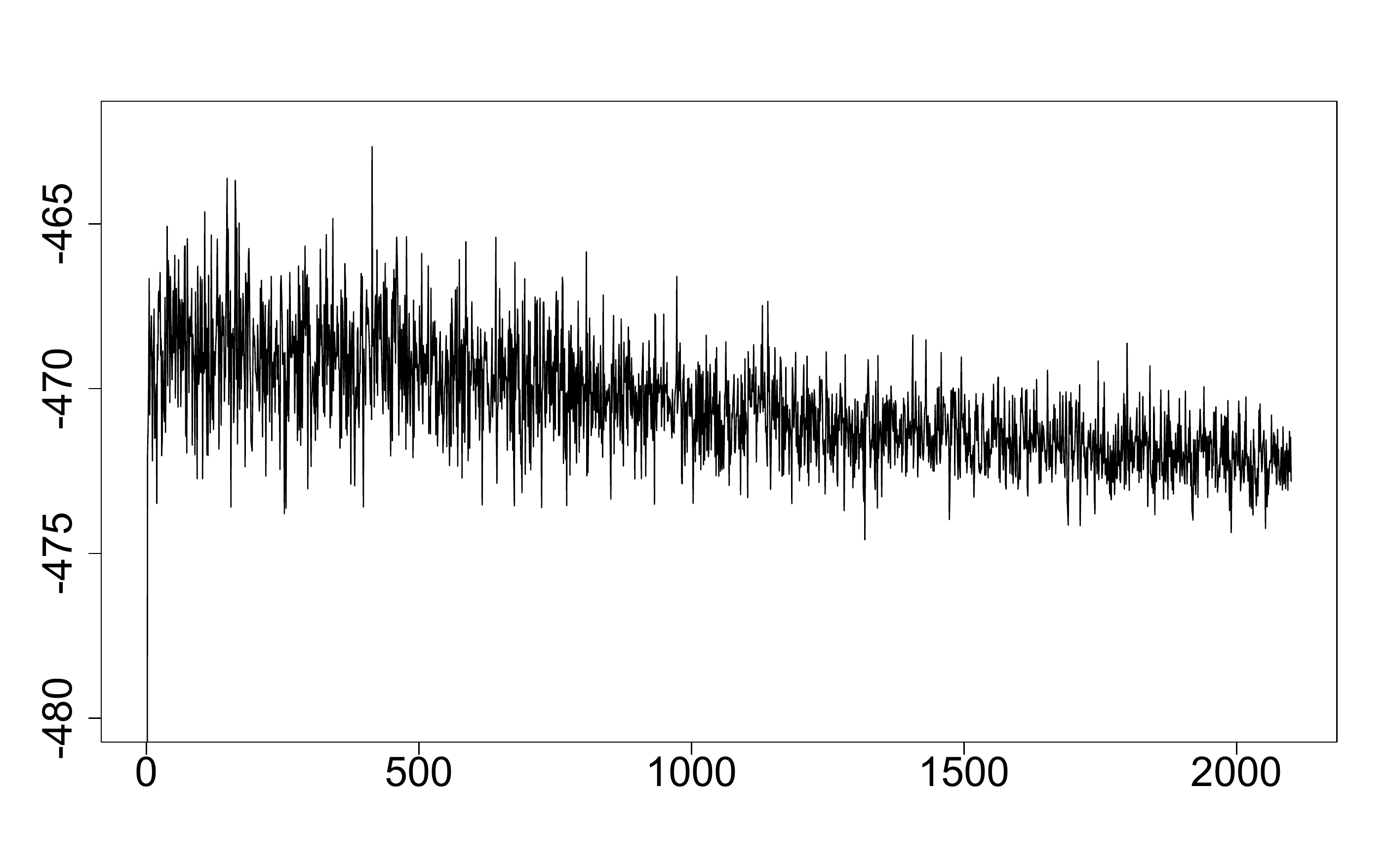}};  
    \node[below=of img8,yshift=1cm] (img9)  {\includegraphics[scale=0.23]{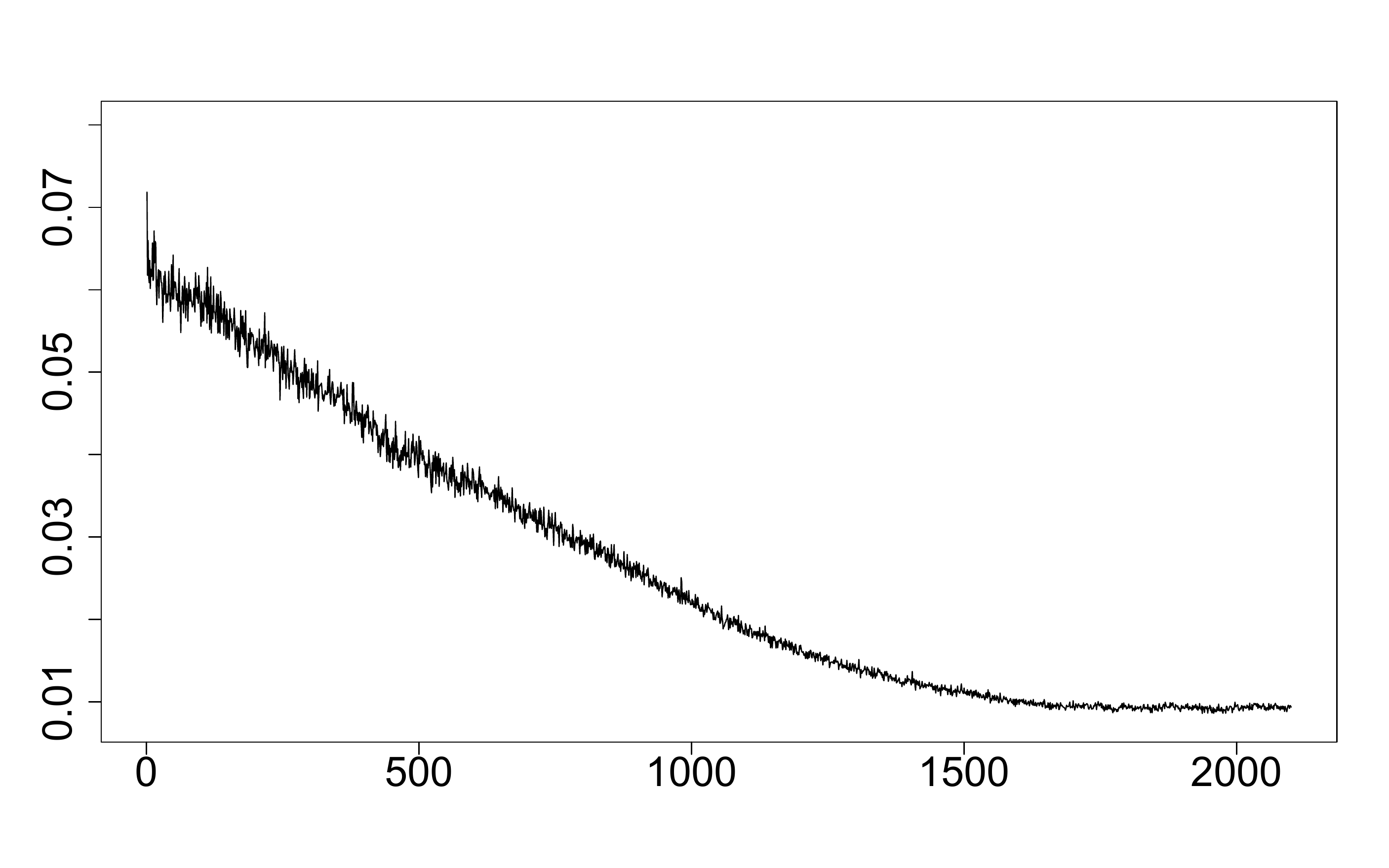}};  
    \node[below=of img9,yshift=1cm] (img10)  {\includegraphics[scale=0.23]{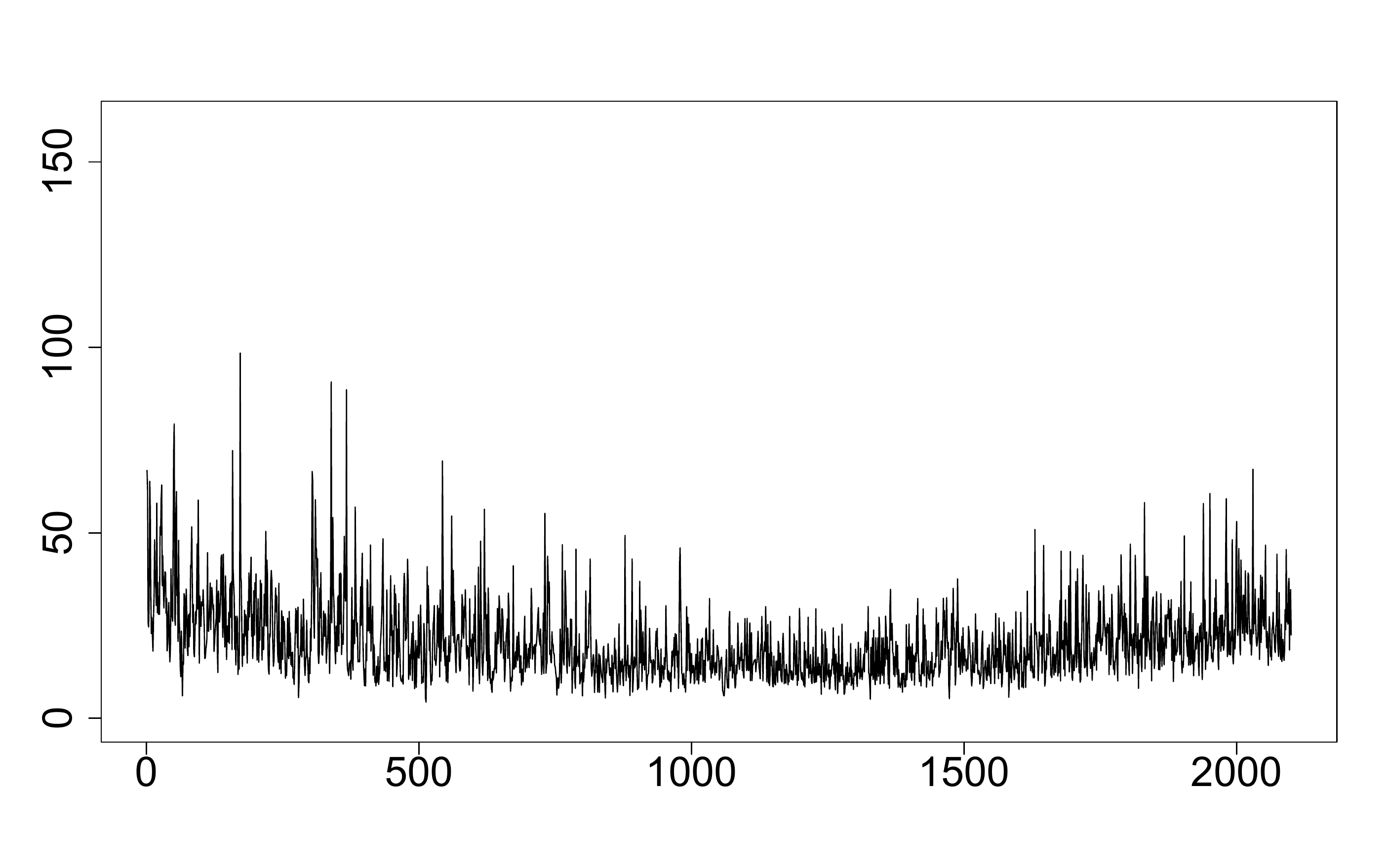}}; 
    forgotten 
  \node[left=of img1, node distance=0cm,  anchor=center,xshift=1cm,font=\small, rotate =0] {$\alpha_{51}$};
  \node[left=of img2, node distance=0cm,  anchor=center,xshift=1cm,font=\small, rotate =0] {$\beta_{51}$};
  \node[left=of img3, node distance=0cm,  anchor=center,xshift=1cm,font=\small, rotate =0] {$\kappa_5$};
  \node[left=of img4, node distance=0cm,  anchor=center,xshift=1cm,font=\small, rotate =0] {$\sigma_{\varepsilon}^2$};
  \node[left=of img5, node distance=0cm,  anchor=center,xshift=1cm,font=\small, rotate =0] {$\sigma_{\omega}^2$};
    
  \node[above=of img1, node distance=0cm, yshift=-1.4cm,font=\small] {Proposed Initial Values};
  
  \node[above=of img6, node distance=0cm, yshift=-1.4cm,font=\small] {SVD Initial Values};
\end{tikzpicture}
\caption{The trace plots of five selected parameters $\alpha_{51}$ (age 50), $\beta_{51}$ (age 50), $\kappa_5$ (year 2000), $\sigma_{\varepsilon}^2$, and $\sigma_\omega^2$. The left column represents 2,100 iterations (including 100 burn-ins) started with the proposed initials values while the right column uses the SVD estimates as the initial values.}
\label{fig:3}
\end{figure}
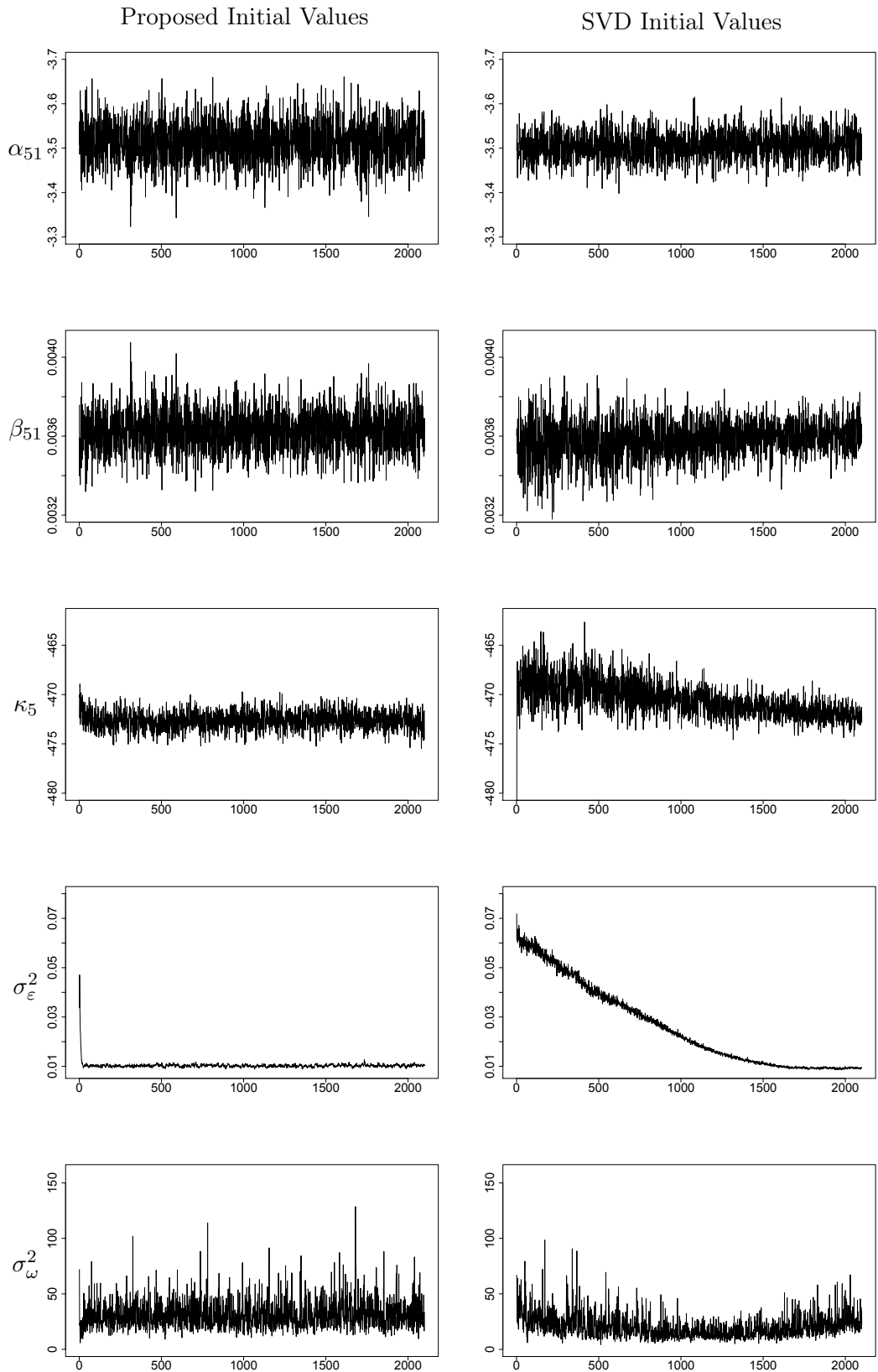

Under the dirac spike setting, the MCMC chains switch back and forth between two time effect models. Therefore, it is required to determine the structure of (\ref{eq3.3}) before any posterior inferences or predictions. Here, we select a random walk with drift model for $\kappa_t$ since there are 1,881 out of 2,000 iterations (around 94\%) with $\theta_2=0$. Accordingly, based on these 1,881 iterations, we compute the posterior means of $\alpha_x$'s, $\beta_x$'s, and $\kappa_t$'s in Figure 4 (left column), and construct the mortality projections for the years 2017-2040 in Figure 5  (left column) by sampling the posterior predictive distributions of $\kappa_t$ and $\log(\mu_{x,t})$.
Specifically, we obtain an MCMC sample of the future mortality rates via
\begin{align*}
    &\kappa_t^{(j)}\sim N(\kappa_{t-1}^{(j)}+\theta_1^{(j)}, (\sigma_{\omega}^2)^{(j)})
    \end{align*}
    and
    \begin{align*}
    &\log(\mu_{x,t}^{(j)})\sim N(\alpha_x^{(j)}+\beta_x^{(j)}\kappa_t^{(j)}, (\sigma_{\varepsilon}^2)^{(j)}),
\end{align*}
where $j=1,2,\dots,1881$ corresponds to those iterations with $\theta_2=0$, and $t=23,24,\dots,$ $46$ denotes the years 2017-2040. 
It is noticeable that the estimated posterior distributions of $\alpha_x$'s, $\beta_x$'s, and $\kappa_t$'s are all concentrated but not smoothing, implying the proposed sampling algorithm converges well even with this challenging data set. As for the mortality projections, we observe that the 95\% HPD intervals can overall capture the observed mortality rates in these five selected age groups, and that 
the rates drop dramatically from the age groups 0 to 20, but continue gradual increments afterward.  
To echo our speculation regarding to the missing mechanism, we also include the results based on the imputed complete data set in Figures 4 and 5 (right column). 
Besides the same time structure is selected (1,865 out of 2,000 iterations with $\theta_2=0$), the complete data set yields the comparable estimations and predictions.

\begin{figure}[H]
\centering
\begin{tikzpicture}
  \node (img1)  {\includegraphics[scale=0.25]{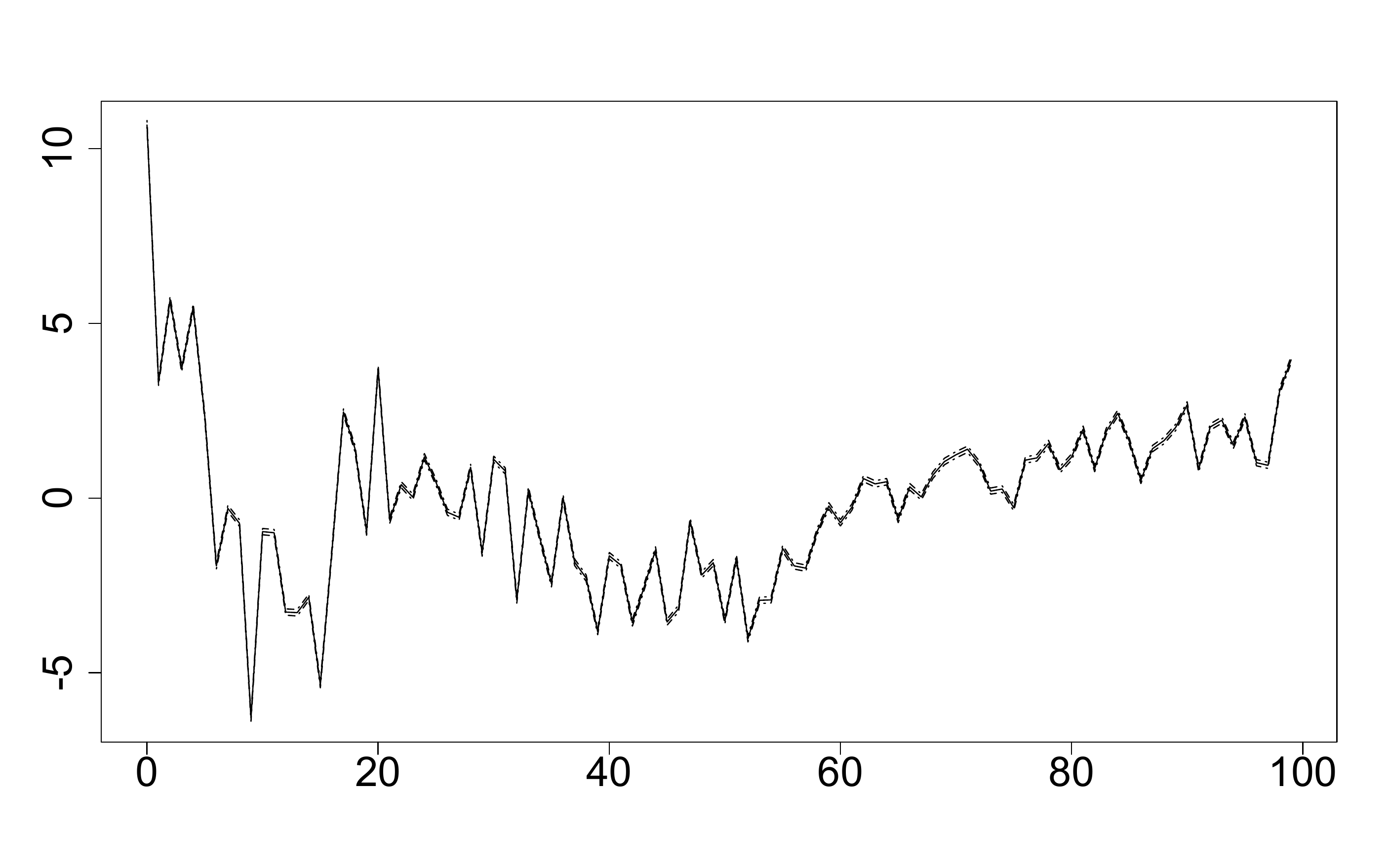}};
  \node[left=of img1, node distance=0cm,  anchor=center,xshift=1cm,font=\small] {$\alpha_x$};
  \node[below=of img1, node distance=0cm,  xshift=0cm, yshift=1.6cm,font=\small] {Age};
  
  \node[right=of img1, xshift=-1cm] (img2)  {\includegraphics[scale=0.25]{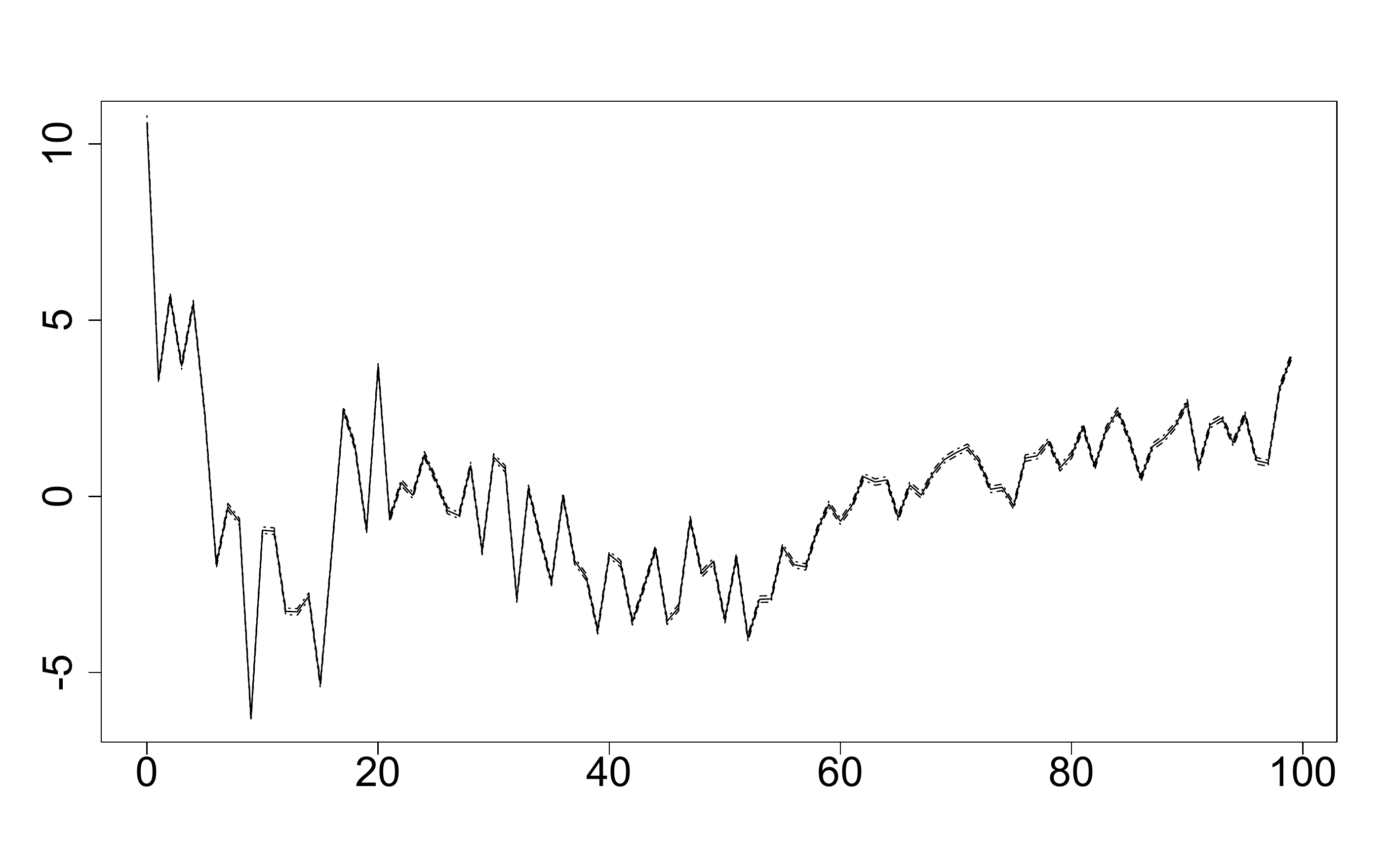}};
  \node[below=of img2, node distance=0cm,  xshift=0cm, yshift=1.6cm,font=\small] {Age};

  \node[below=of img1, yshift=0.5cm] (img3)  {\includegraphics[scale=0.25]{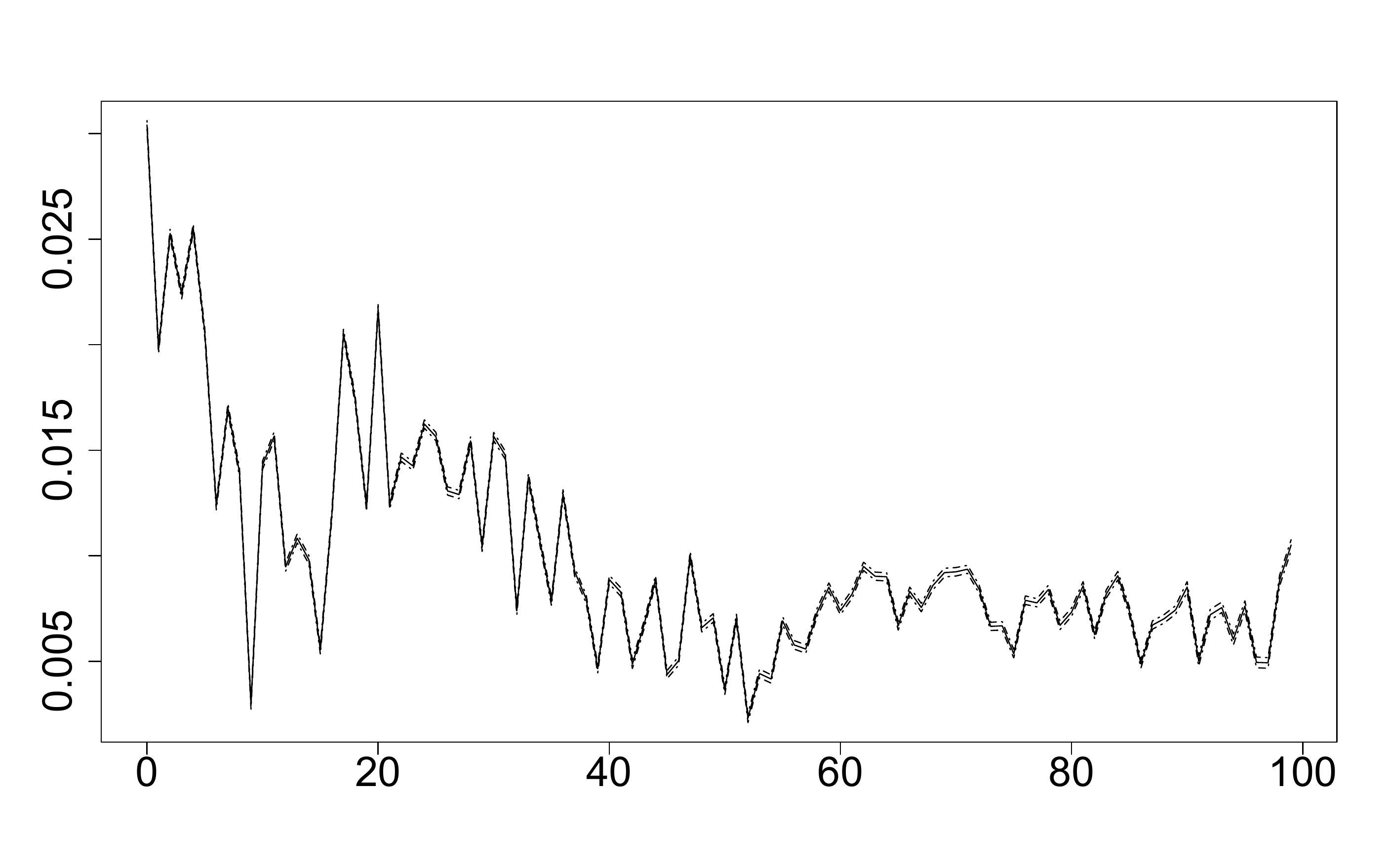}};
  \node[left=of img3, node distance=0cm,  anchor=center,xshift=1cm,font=\small] {$\beta_x$};
  \node[below=of img3, node distance=0cm,  xshift=0cm, yshift=1.6cm,font=\small] {Age};
   
  \node[below=of img2,yshift=0.5cm] (img4)  {\includegraphics[scale=0.25]{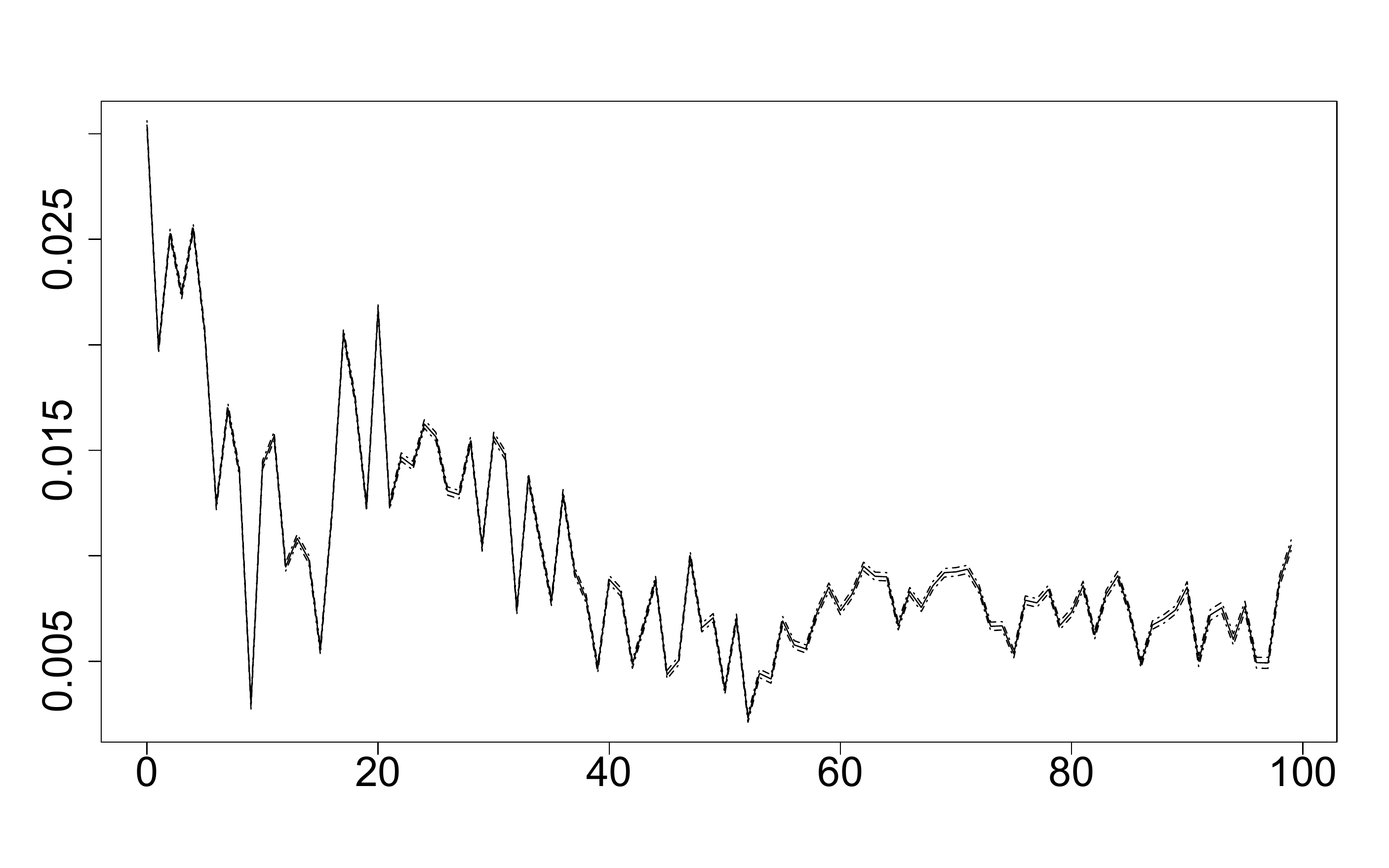}};
  \node[below=of img4, node distance=0cm,  xshift=0cm, yshift=1.6cm,font=\small] {Age};
  
  \node[below=of img3, yshift=0.5cm] (img5)  {\includegraphics[scale=0.25]{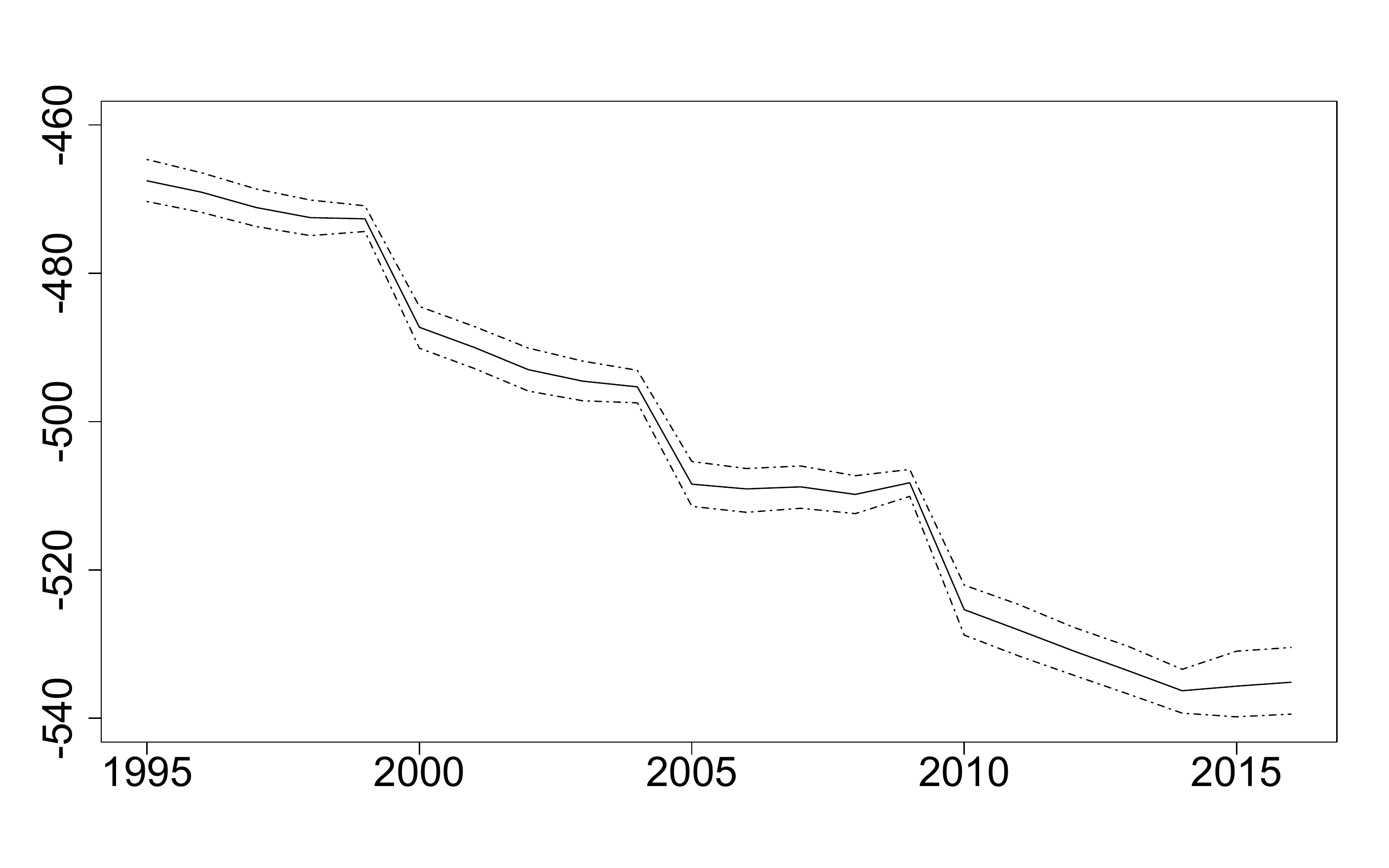}};
  \node[left=of img5, node distance=0cm,  anchor=center,xshift=1cm,font=\small] {$\kappa_t$};
  \node[below=of img5, node distance=0cm,  xshift=0cm, yshift=1.6cm,font=\small] {Year};
  
  \node[right=of img5, xshift=-1cm] (img6)  {\includegraphics[scale=0.25]{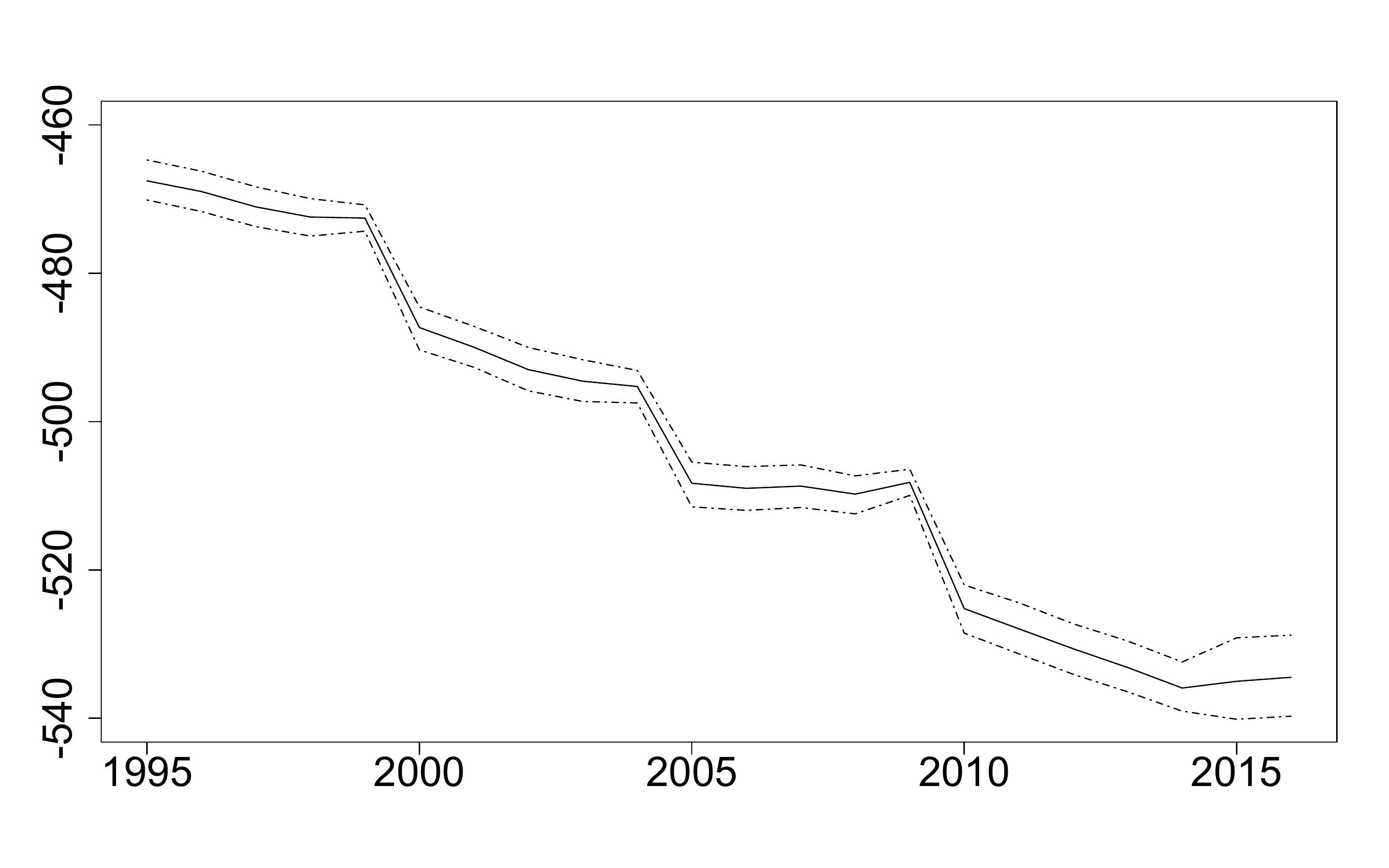}};
  \node[below=of img6, node distance=0cm,  xshift=0cm, yshift=1.6cm,font=\small] {Year};
  
  \node[above=of img1, node distance=0cm, yshift=-1.4cm,font=\small] {Incomplete Data};
  \node[above=of img2, node distance=0cm, yshift=-1.4cm,font=\small] {SVD-imputed Complete Data};
\end{tikzpicture}
\caption{The posterior means of $\alpha_{x}$'s, $\beta_{x}$'s, and $\kappa_t$'s along with 95\% HPD intervals (dash-dotted lines). The left column presents the results of the proposed model based on incomplete data while the right one is based on SVD-imputed complete data. }
    \label{fig:7}
    
\end{figure}
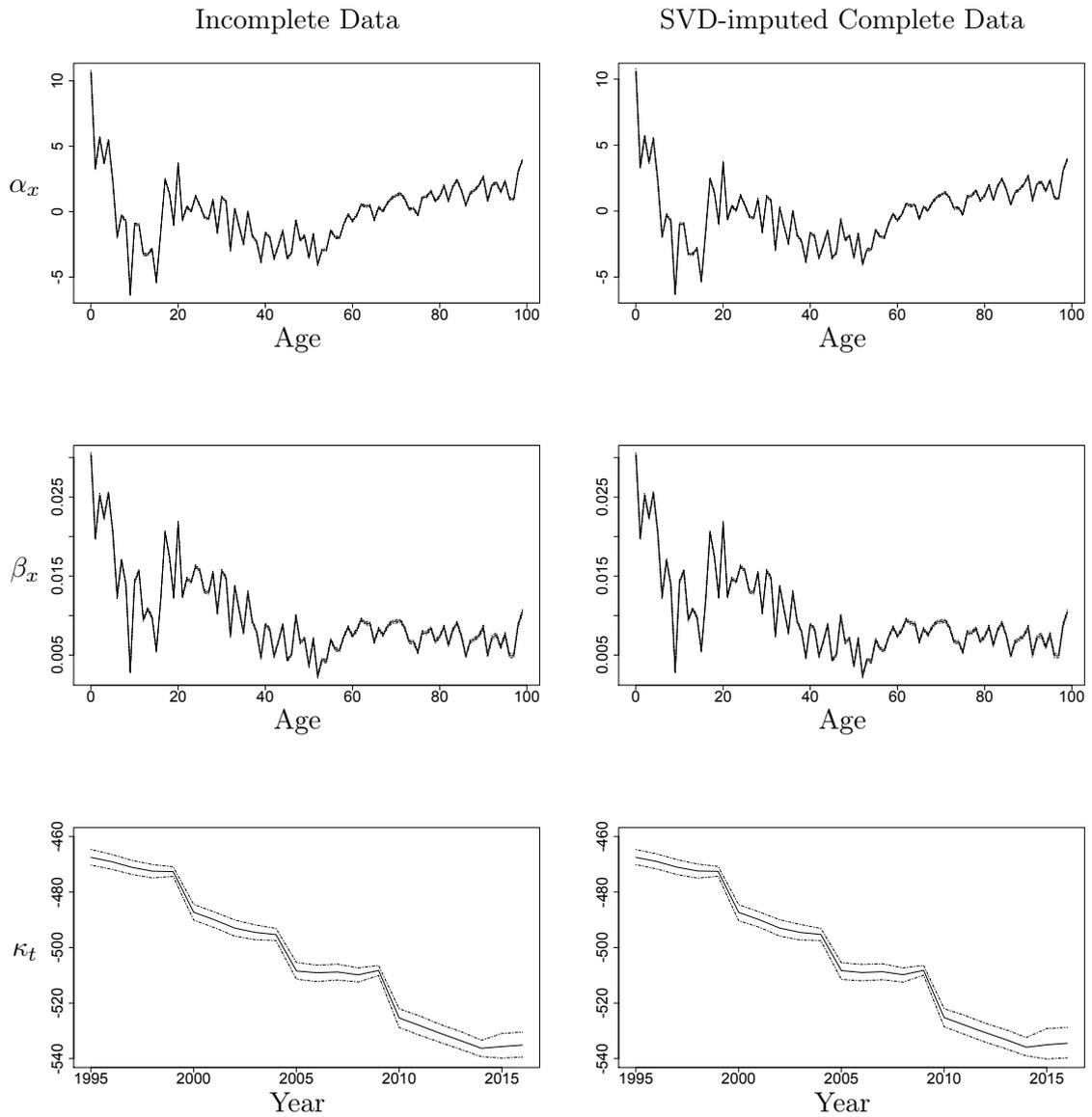







\begin{figure}[H]
\centering
\begin{tikzpicture}
  \node (img1)  {\includegraphics[scale=0.2]{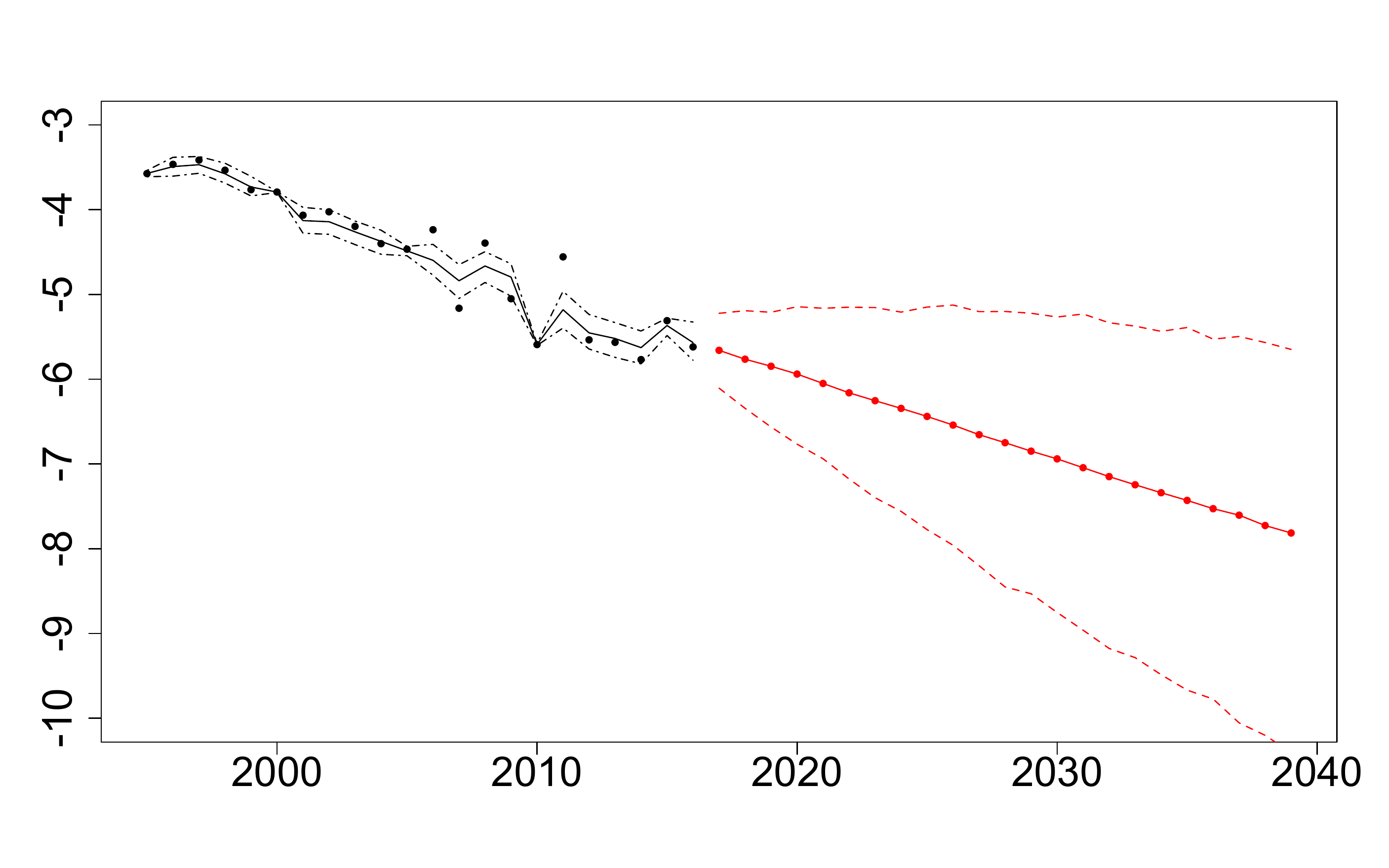}};
  \node[left=of img1, node distance=0cm, xshift=1cm,font=\small] {Age 0};
  \node[below=of img1, yshift=0.5cm] (img2)  {\includegraphics[scale=0.2]{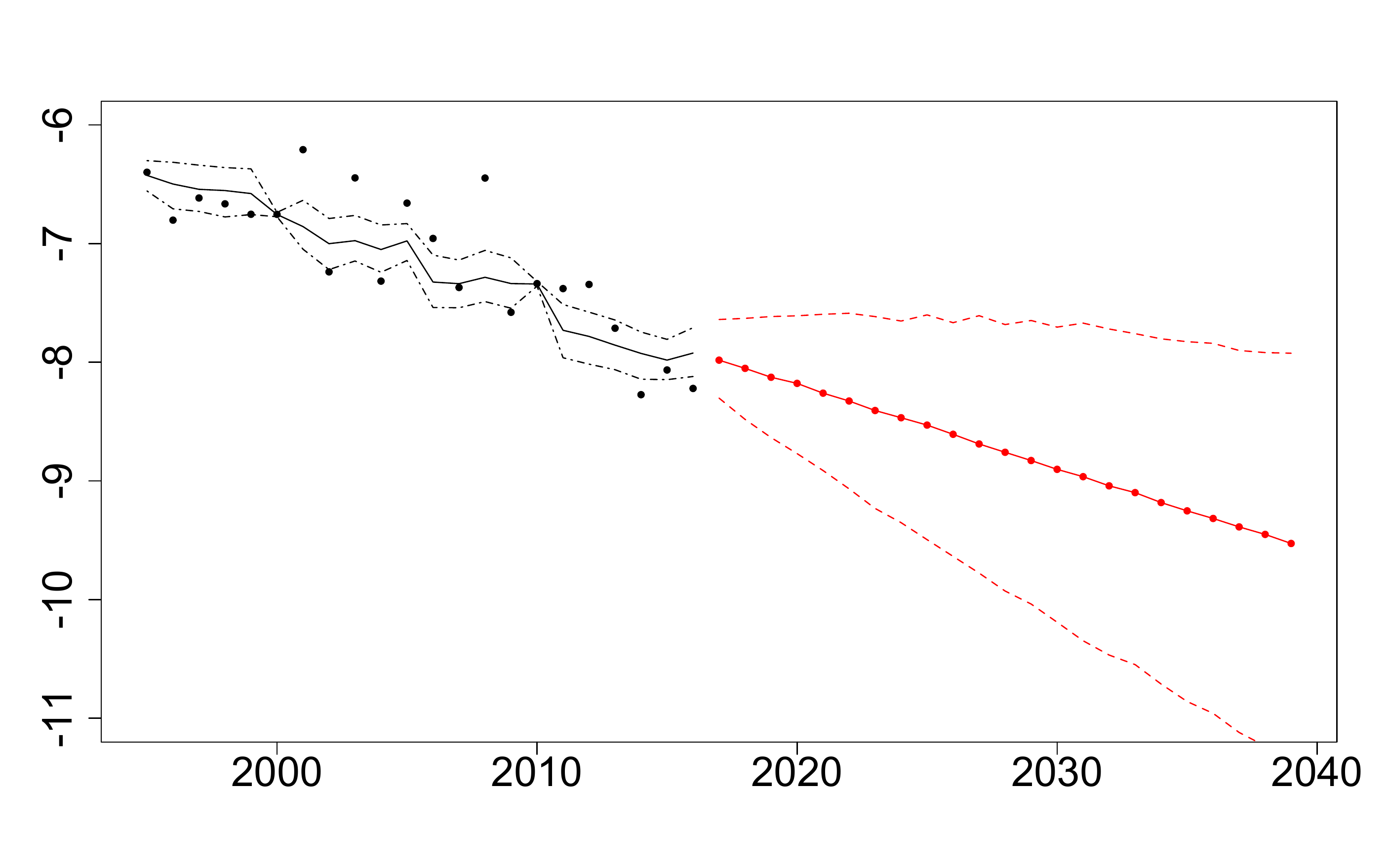}};
  \node[below=of img2, yshift=0.5cm] (img3)  {\includegraphics[scale=0.2]{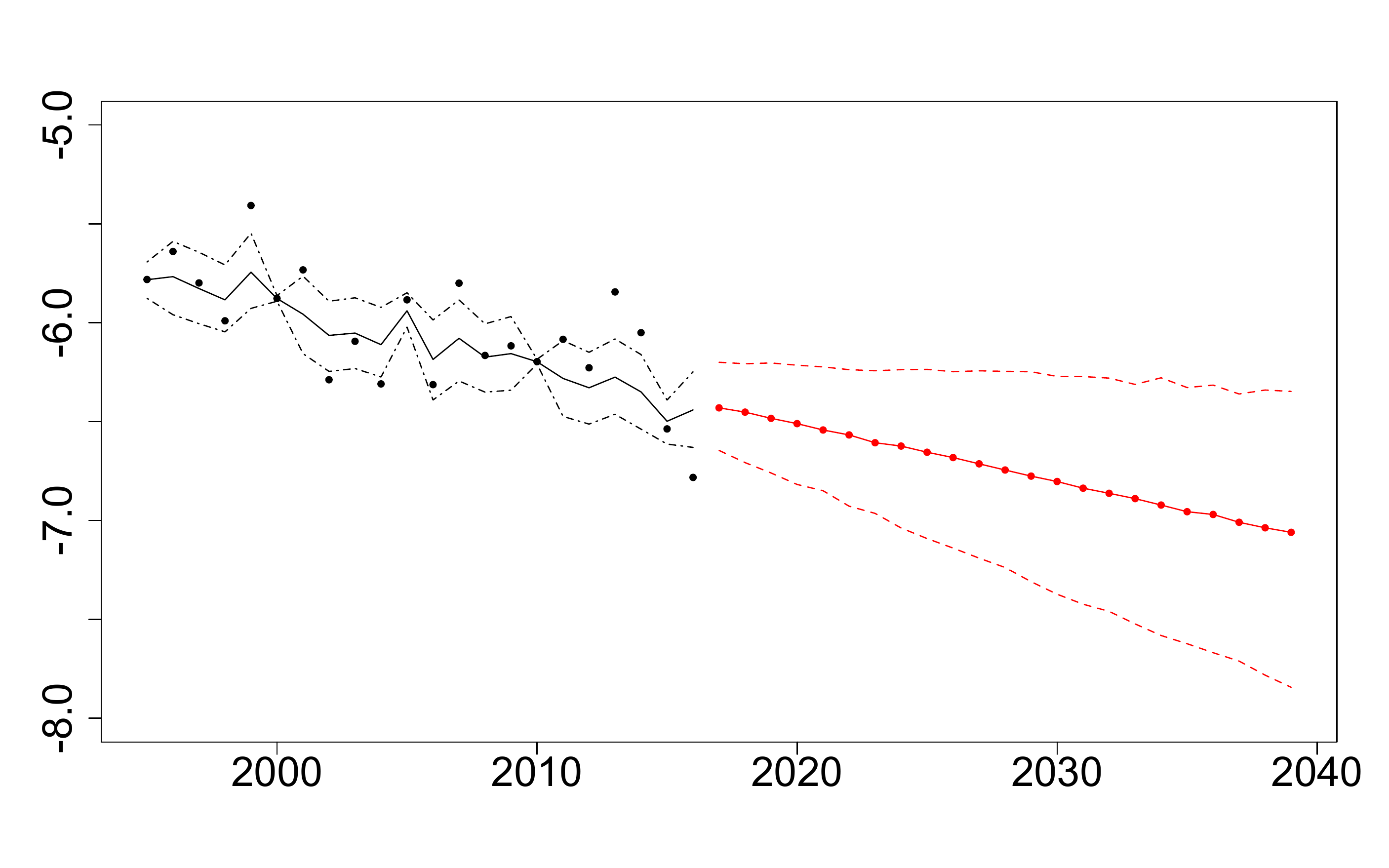}};
    \node[below=of img3, yshift=0.5cm] (img4)  {\includegraphics[scale=0.2]{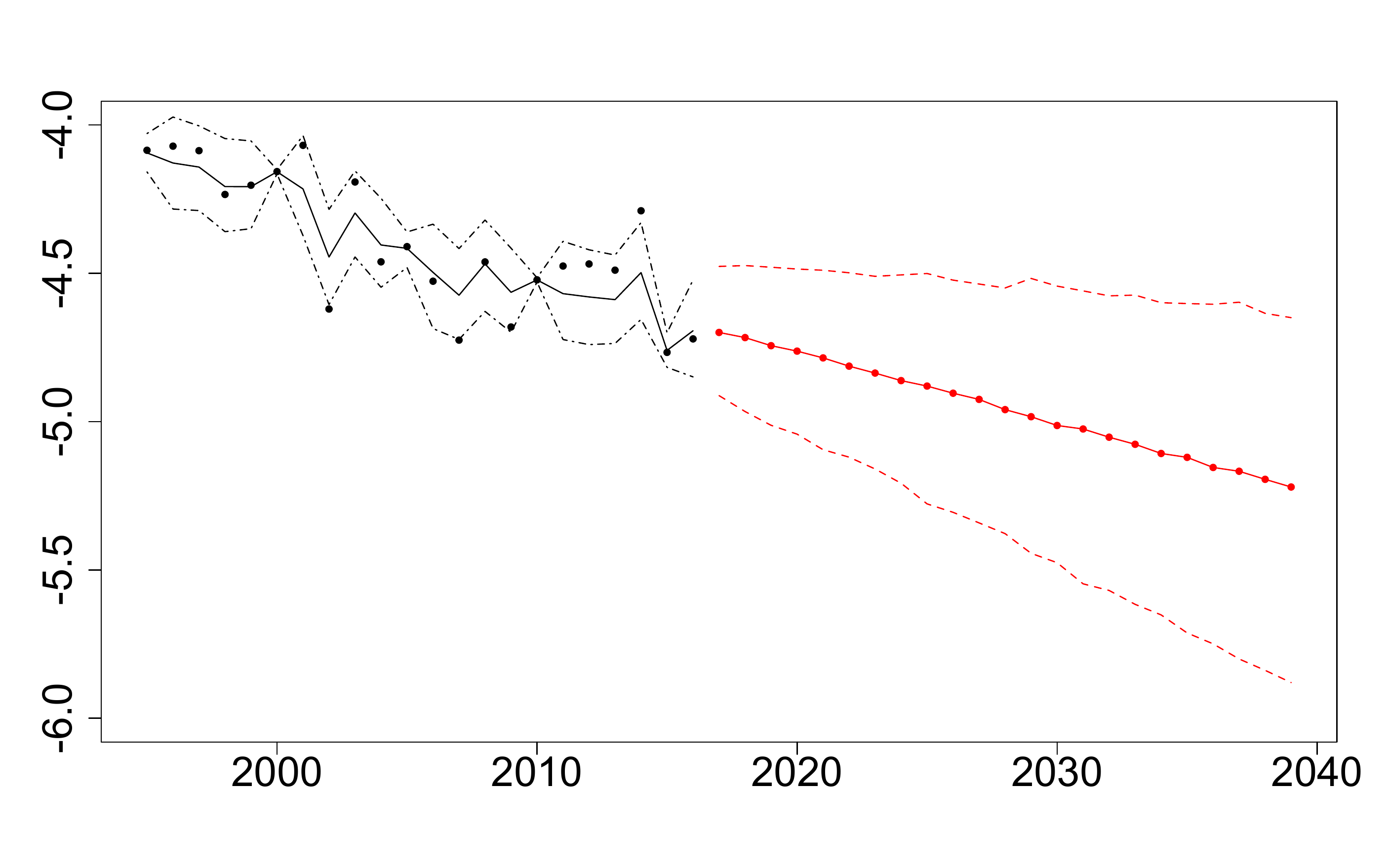}};
    \node[below=of img4, yshift=0.5cm] (img5)  {\includegraphics[scale=0.2]{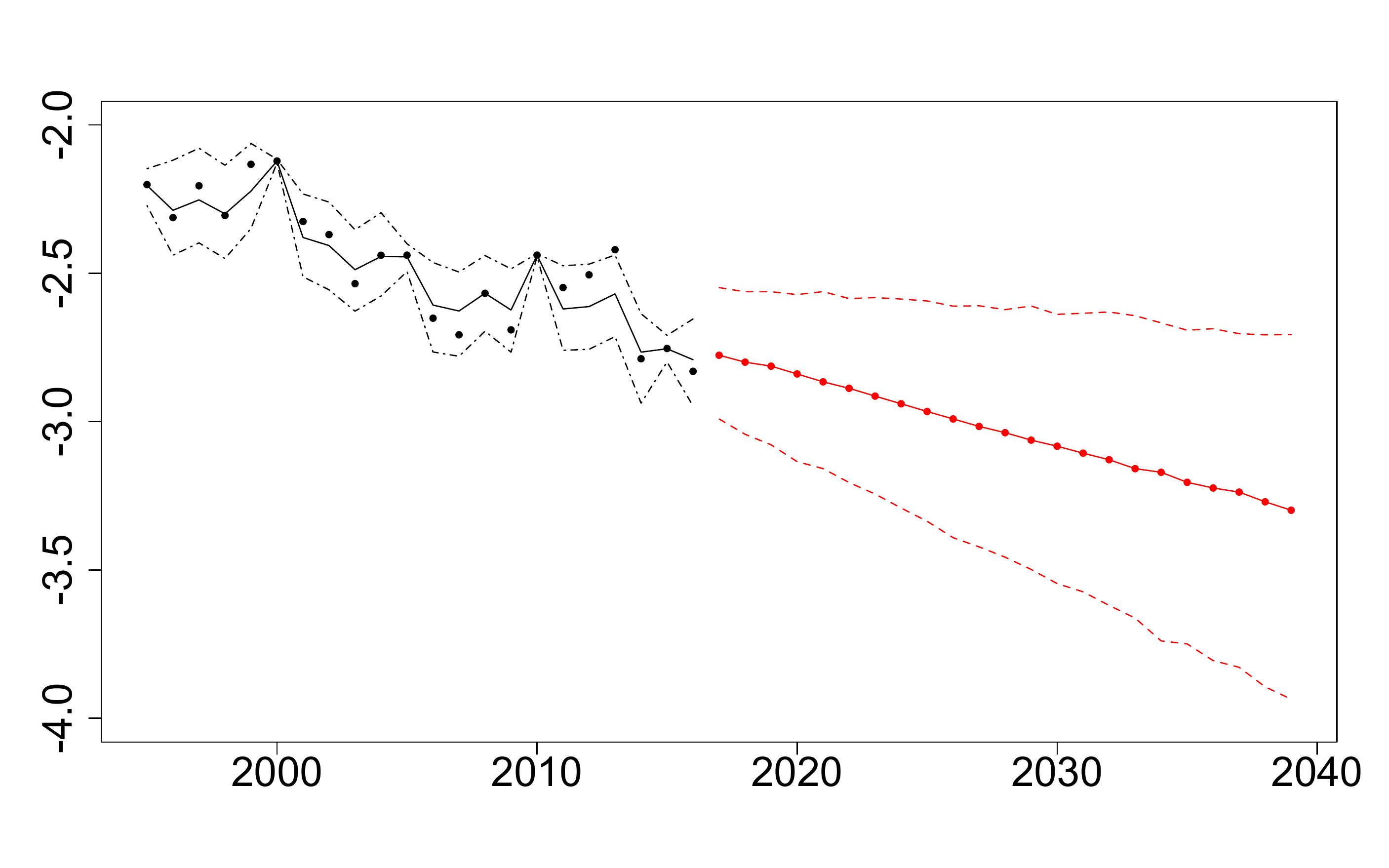}};
    
    \node[right=of img1, xshift=-1cm] (img6)  {\includegraphics[scale=0.2]{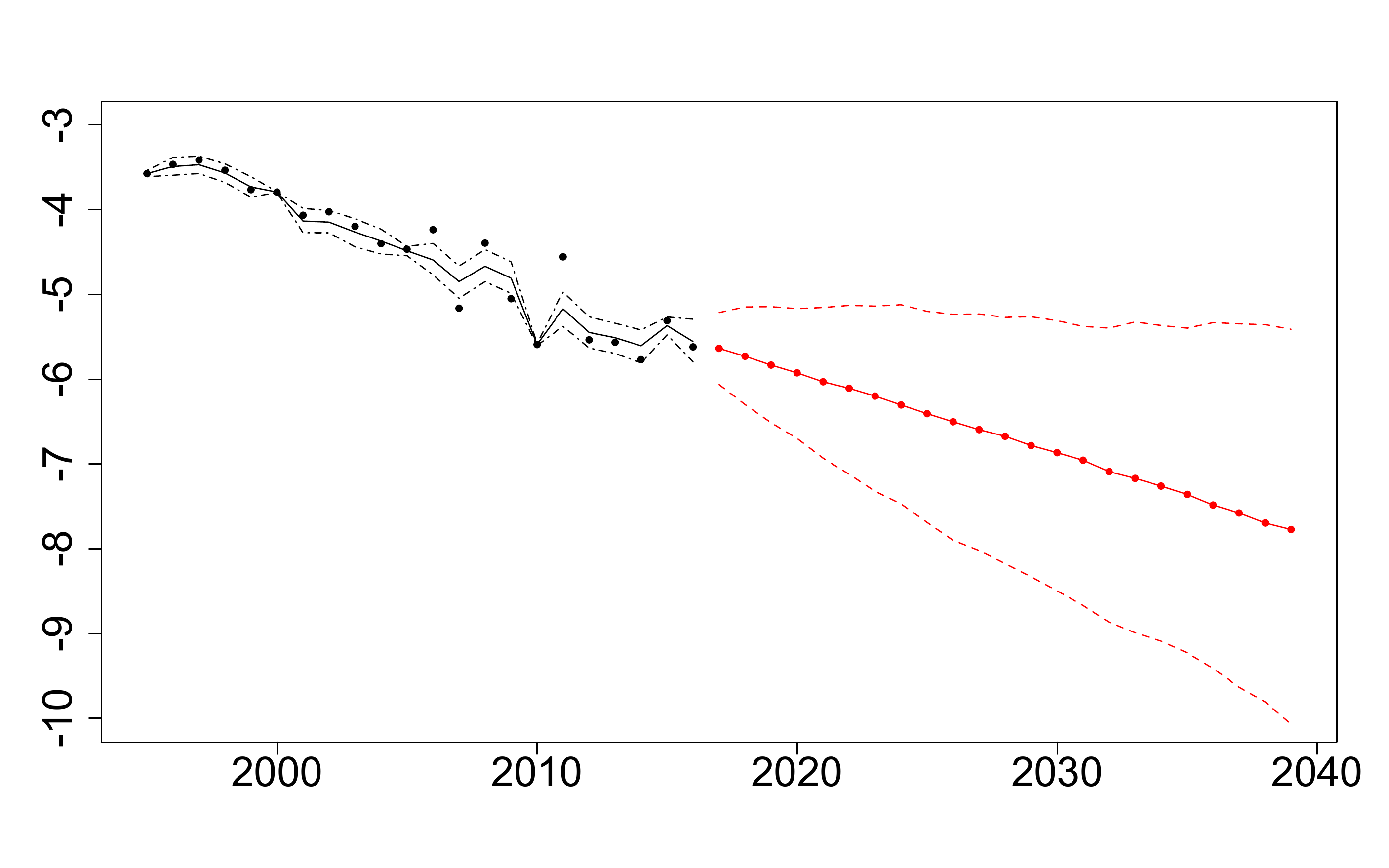}};
    \node[below=of img6, yshift=0.5cm] (img7)  {\includegraphics[scale=0.2]{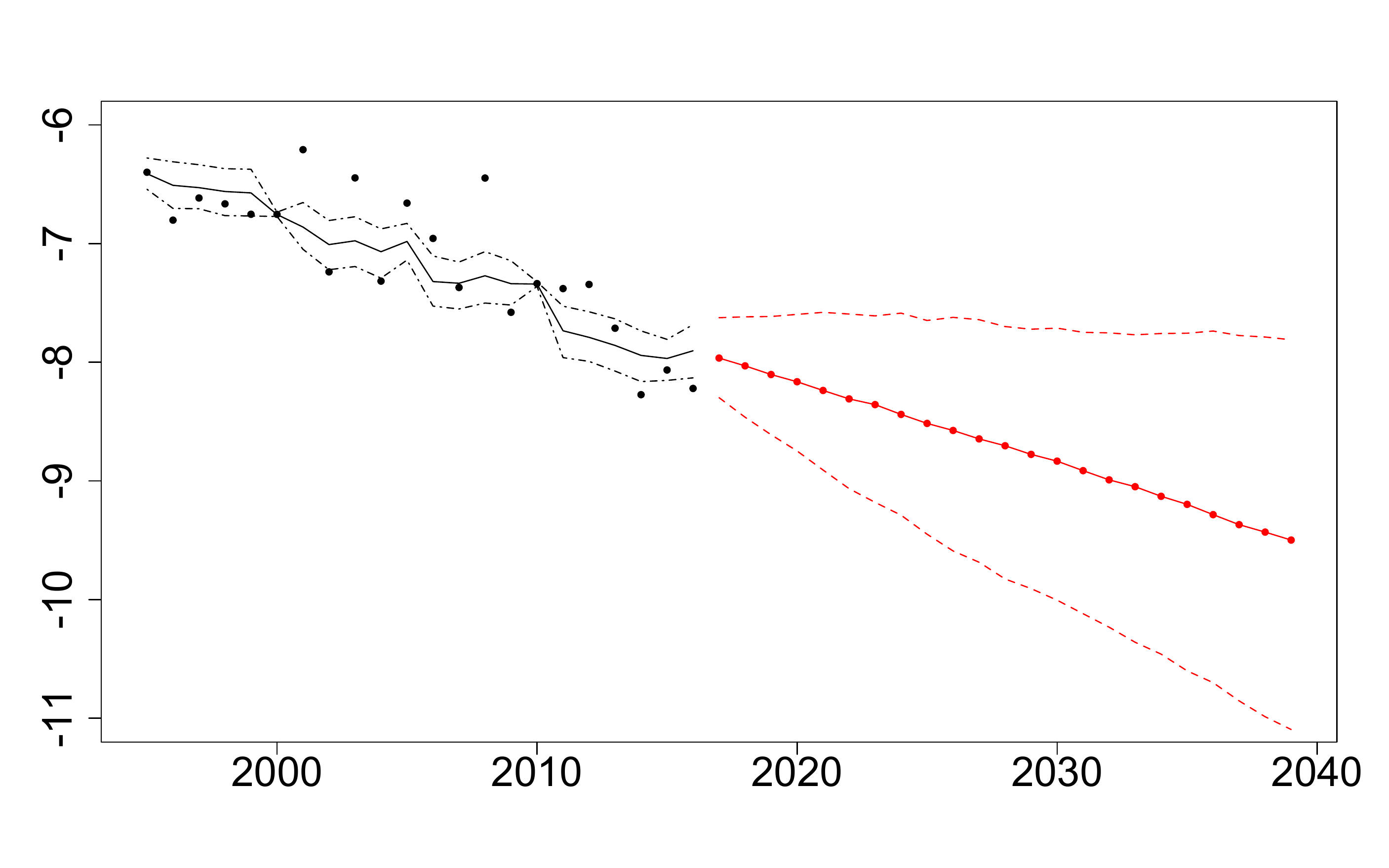}};
    \node[below=of img7, yshift=0.5cm] (img8)  {\includegraphics[scale=0.2]{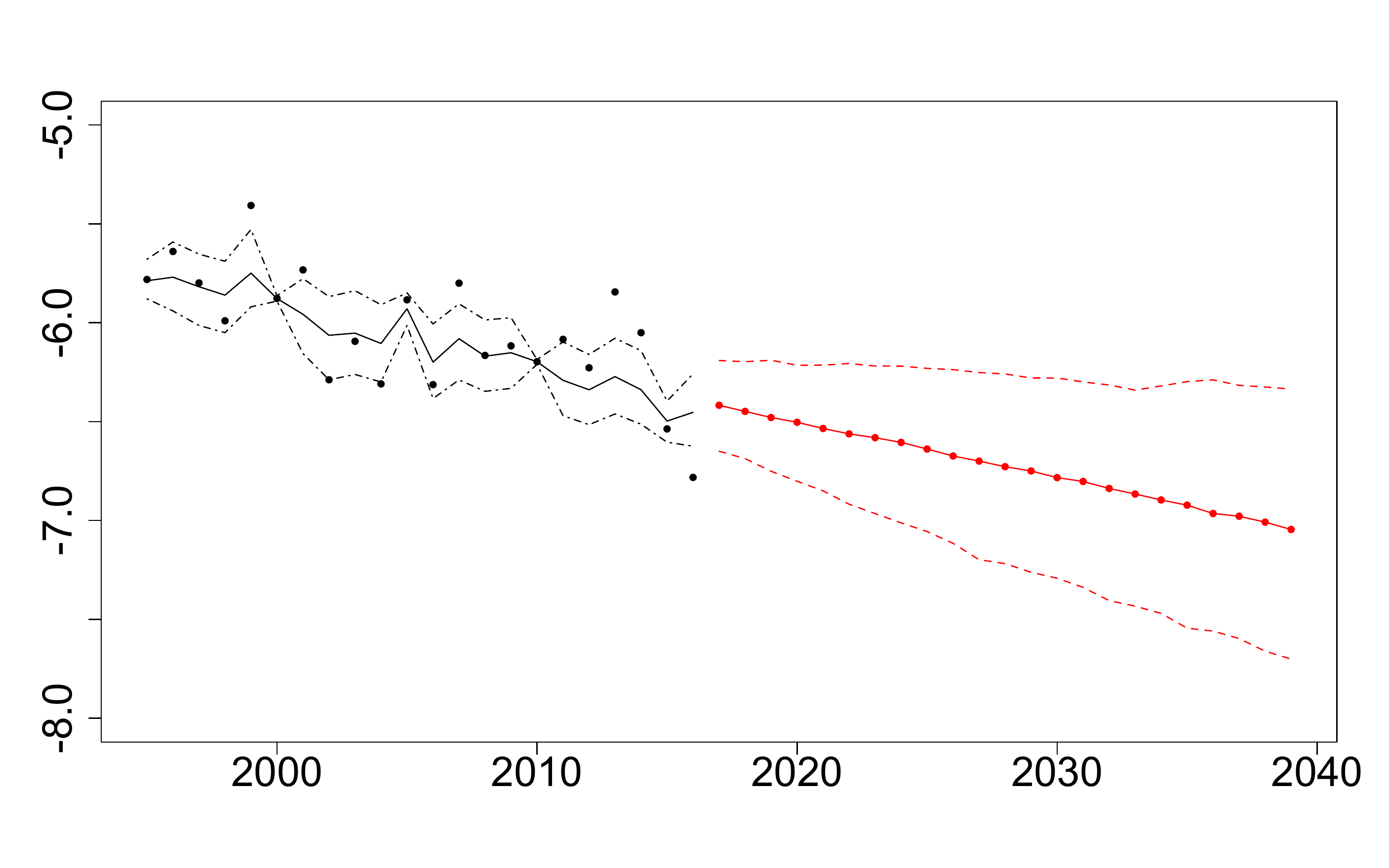}};
    \node[below=of img8, yshift=0.5cm] (img9)  {\includegraphics[scale=0.2]{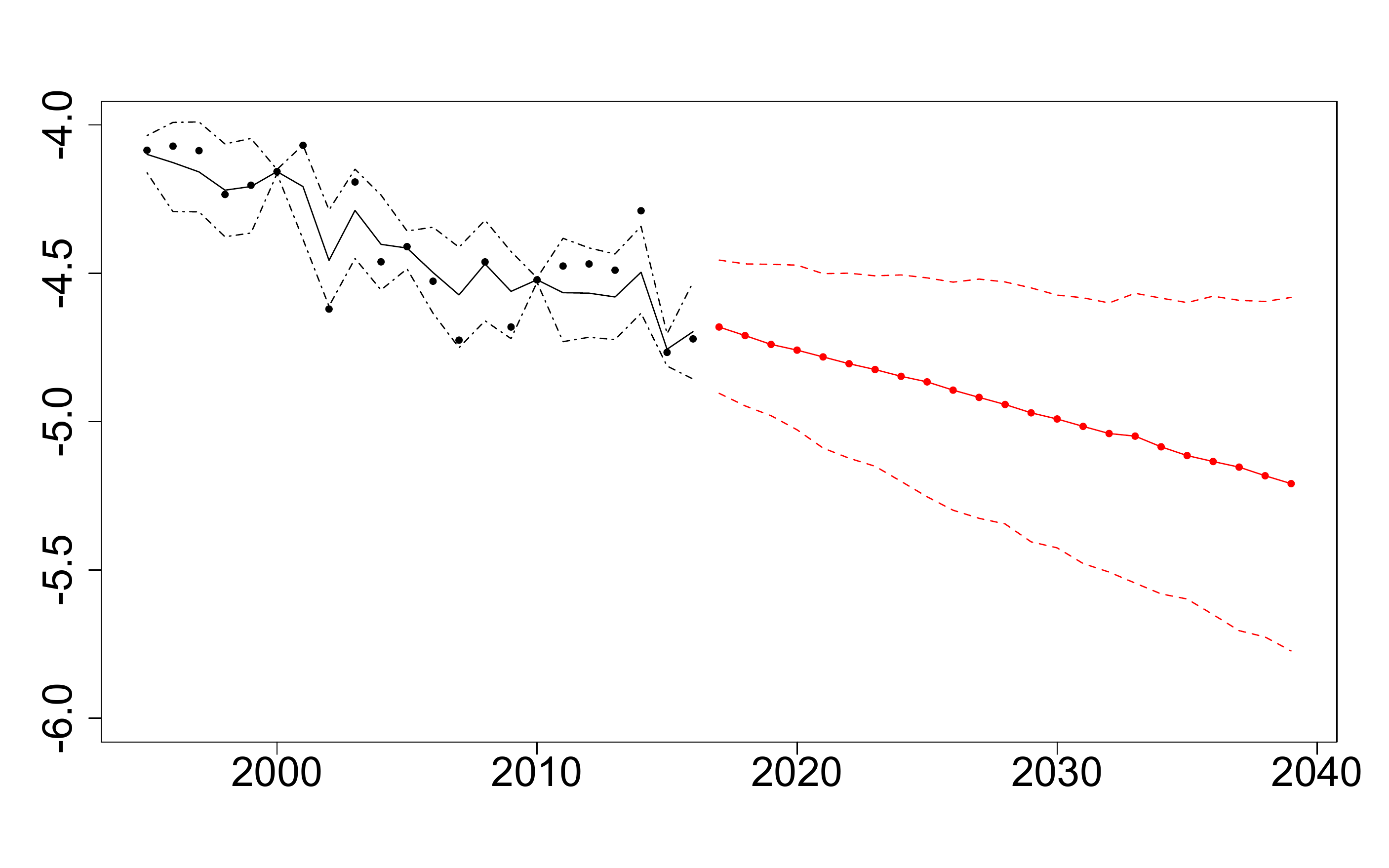}};
    \node[below=of img9, yshift=0.5cm] (img10)  {\includegraphics[scale=0.2]{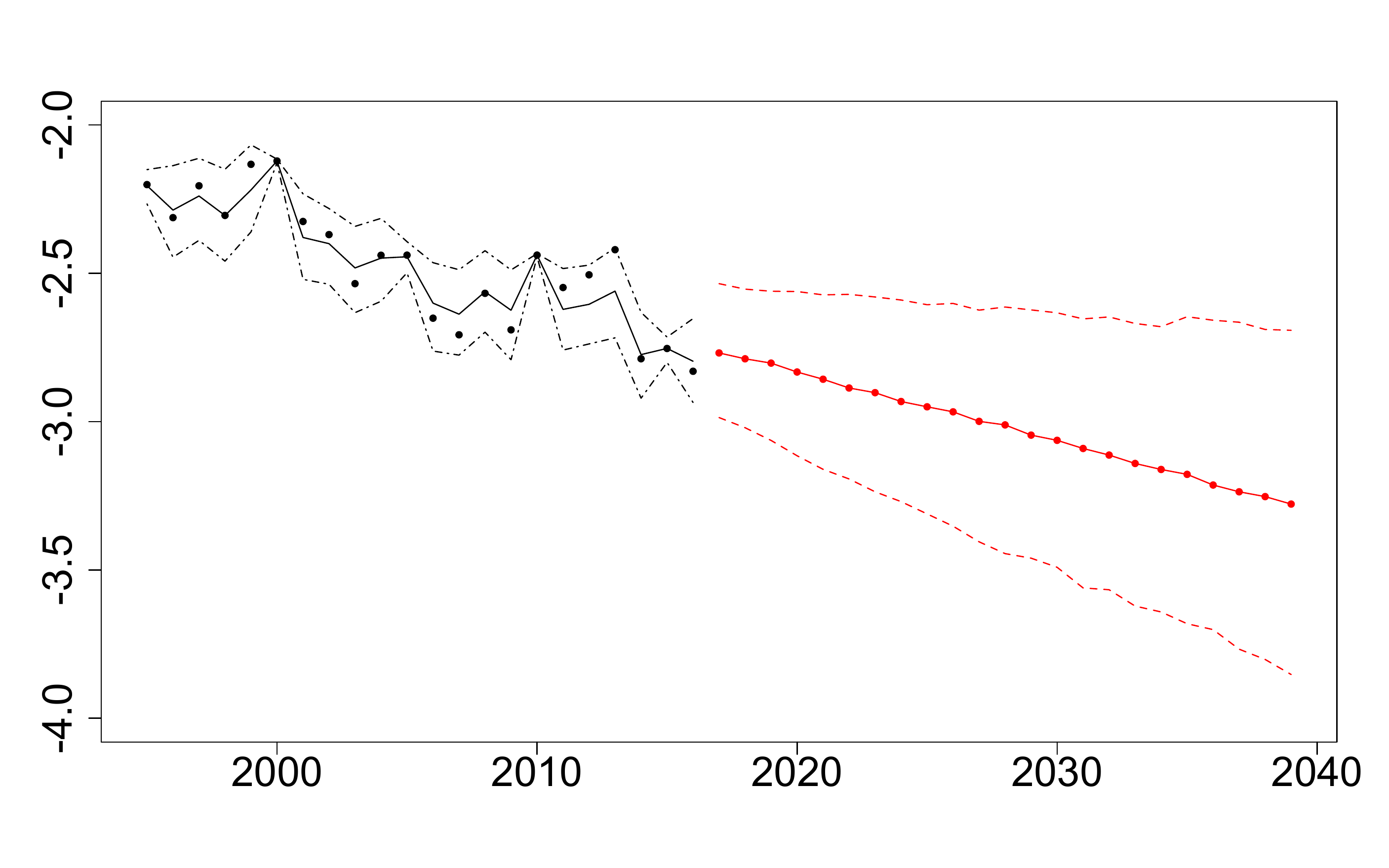}};
    
  \node[left=of img2, node distance=0cm, xshift=1cm,font=\small] {Age 20};
  \node[left=of img3, node distance=0cm, xshift=1cm,font=\small] {Age 40};
  \node[left=of img4, node distance=0cm, xshift=1cm,font=\small] {Age 60};
  \node[left=of img5, node distance=0cm, xshift=1cm,font=\small] {Age 80};
  
    \node[above=of img1, node distance=0cm, yshift=-1.4cm,font=\small] {Incomplete Data};
  \node[above=of img6, node distance=0cm, yshift=-1.4cm,font=\small] {SVD-imputed Complete Data};
\end{tikzpicture}
\caption{The mortality projections of selected age  groups: 0, 20, 40, 60, and 80 years old, along with 95\% HPD intervals (dash-dotted lines), where the black dots denote the observed mortality rates, and the red ones are predicted mortality rates. The left column presents the results of the proposed model based on incomplete data while the right one is based on SVD-imputed complete data.}
    \label{fig:7}
    
\end{figure}
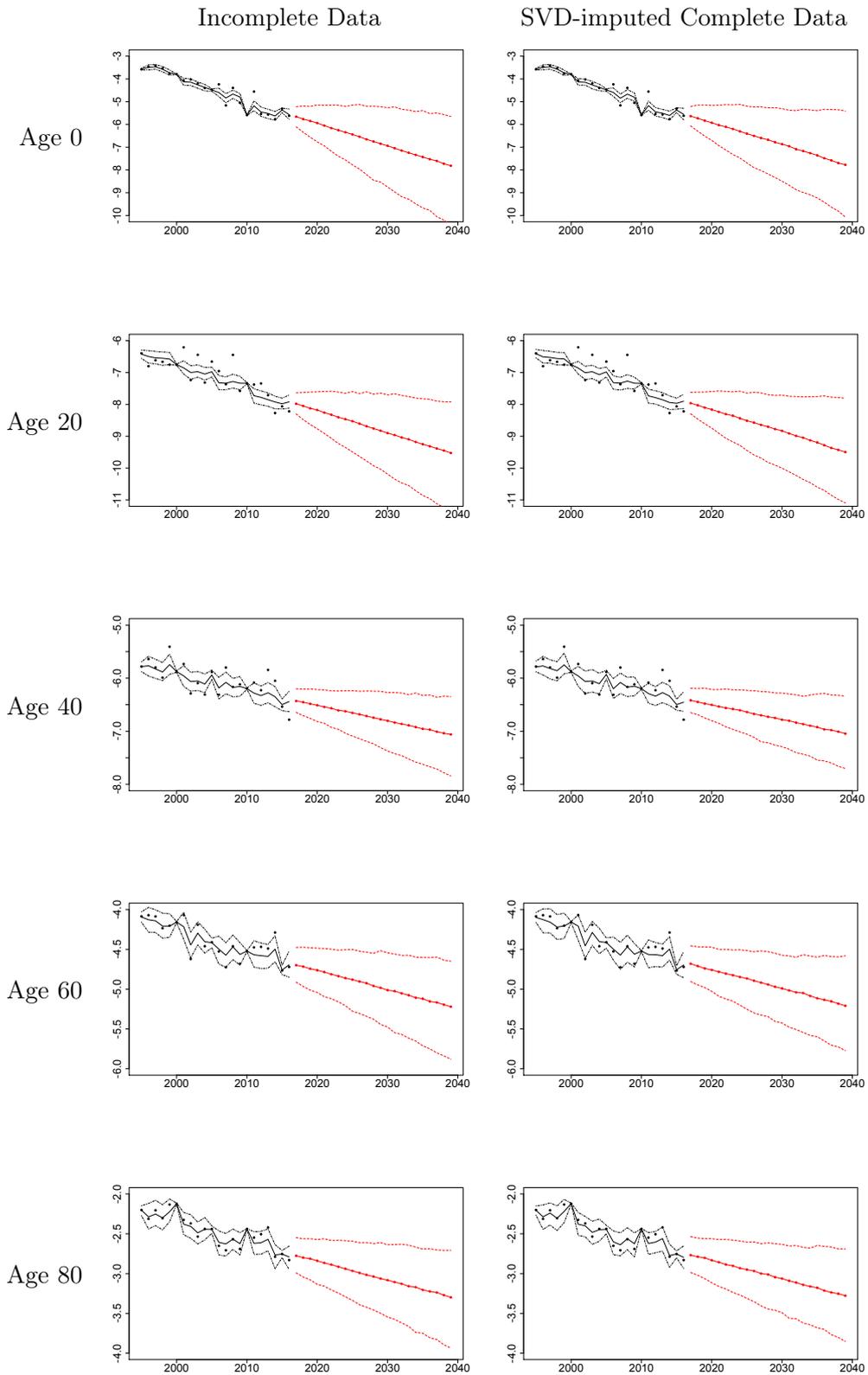

%% file: Section6.tex
\section{Conclusion}

\hspace{12pt} In this work, we present the extended PLNLC model along with the new MCMC sampling algorithm, where a more flexible setting of the time structure is considered while the selection between the full and reduced structures can be done simultaneously with estimations and predictions. By combining the Kalman and sequential Kalman filters into the Gibbs sampling, the proposed model can efficiently update $\kappa_t$ even when applied to a challenging data set such as with missing observations or dramatic changes in the mortality rates of two adjacent ages or years. Since this algorithm only requires log mortality rates, which can be easily obtained in the Gibbs sampler, to claim as the state space form, and these rates are also the outputs commonly used in evaluating the goodness of fit, our proposed sampling algorithm adds at no additional computational cost. Besides, with the constraints \citet{Johnny} embedded into the prior specifications, 
we have the time effect well interpreted as an aggregation of log mortality rate in that year, and avoid any potential violations of ergodic conditions in the sampling scheme.

We also view this work as a twins of \citet{Johnny}, and fill the gap in mortality modelling under the Poisson framework. Additionally, via the dirac spike setting, the proposed approach and sampling algorithm can be easily adjusted to accommodate a more complex time effect structure while does not erase the possibility of being simple, as long as the full model can be expressed as the state space form. 
It is also worth pointing out that the sequential update of $\kappa_t$ is particularly desirable when the large administrative region is of interest because its missing observations could be attributed to few subregions failure to provide the death counts. Although this issue is commonly addressed by utilizing information from other subregions to retain the estimated total deaths, this approach ignores the uncertainty from the subregion level. Alternatively, we can extend the sequential Kalman filter to allow the updates directly based on the subregion mortality data, that is, we now have the recursive equations designed for the age-year-and-subregion data. We mark this as a potential future work. 